\documentclass[aps,prr,superscriptaddress, twocolumn]{revtex4-1}

\usepackage[english]{babel}
\usepackage{times}
\usepackage{graphicx}
\usepackage{graphics}
\usepackage{amsmath}
\usepackage{mathtools}
\usepackage{amsfonts}
\usepackage{amssymb}
\usepackage{dsfont}
\usepackage{epstopdf}
\usepackage{makeidx}
\usepackage{subfigure}
\usepackage{color}
\usepackage{pgf}
\usepackage{bm}
\usepackage{ulem}
\usepackage{tikz} 

\makeindex

\begin{document}



\title{{Gap fluctuations, Cooper pairs with finite center-of-mass momentum, and suppression of superconductivity in inhomogeneous systems with dopant superpuddles agglomerates}}

\author{Victor Velasco *,} 

\affiliation{Instituto de Física, Universidade Federal do Rio de Janeiro, Caixa Postal 68528, Rio de Janeiro, Brazil}

\author{Marcello B. Silva Neto} 

\affiliation{Instituto de Física, Universidade Federal do Rio de Janeiro, Caixa Postal 68528, Rio de Janeiro, Brazil}

\begin{abstract}
{Spatially extended aggregates or clusters of dopants are ubiquitous in a plethora of granular superconducting systems, such as Al-doped $\mathrm{MgB_2}$ and N-doped $\mathrm{Mo_2N}$, forming a droplet network that is very important to their characterization and to the description of their superconducting properties.} At the same time, one of the most studied classes of unconventional superconducting materials are the high-temperature superconductors, where special attention is given to the hole-doped cuprates, where the carrier concentration is controlled by the amount of extra interstitial oxygen dopants. In this context, the formation of spatially inhomogeneous aggregates of interstitial dopant oxygen atoms, in the form of nanosized superpuddles, is not only relevant, but also a subject of intense recent experimental and theoretical surveys.  Following these efforts, in this work we investigate the consequences of the presence of networks of {inhomogeneously distributed dopant} superpuddles on the superconducting state. Starting from the inhomogeneous extended disordered background brought by the network of superpuddles, we demonstrate, with the aid of an effective interaction between electrons mediated by the local vibrational degrees of freedom of each puddle, that the Cooper pairs arising from an attractive interaction in an inhomogeneous medium have a finite center-of-mass momentum, $\mathbf{p}$, that breaks up the Cooper channel. Furthermore, we derive an analytical expression for the amplitude of the superconducting gap, $\Delta_\mathbf{k}$, in terms of disorder and finite center-of-mass momentum and show that amplitude fluctuations are induced in the superconducting state by the presence of the superpuddles, where both the gap and the critical temperature are reduced by disorder and finite momentum pairs. Finally, we discuss our findings in the context of {synchronized networks of superconducting oxygen nano-puddles in cuprates and in other granular superconducting systems. \vspace{0.2cm}\\ * Corresponding author: velasco@if.ufrj.br}
\end{abstract}

\maketitle

\section{Introduction}


Within the Bardeen-Cooper-Schrieffer (BCS) theory of superconductivity, the two quasi-particles forming the bound states that constitute the superconductor, Cooper pairs, have momentum $k$ and $-k$, near the Fermi surface, with oposite spins $\uparrow$ and $\downarrow$, forming a singlet with zero center-of-mass momentum \cite{BCS1957}, in what is usually called the Cooper channel. However, the existence of a finite-momentum superconducting ground state has recently been raised theoretically \cite{Agterberg2020, Wang2015, Chakraborty2019, Wardh2017, Choubey2020, Loder2010} and supported by several experiments in correlated quantum materials \cite{Hamidian2016, Liu2021, Chen2021, Chen2018}. Moreover, the possibility of emergent finite-momentum pair states, in the form of pair density waves, in a variety of well-established superconducting compounds, for example transition-metal dichalcogenides and in cuprates \cite{Edkins2019}, points to the importance of understanding the intrinsic characteristics of these states and their interplay with other common features of these systems, such as disorder \cite{Semenikhin2003} and in the presence of magnetic fields \cite{Annica2022}. 


Although condensed matter models start from the notion of a perfect crystal, a plethora of notable effects are only accessible when this notion is no longer true. One famous example is the problem of the high-$T_c$ superconductivity on cuprates, in which a region of $d$-wave pairing occurs in the form of a dome-shaped area and as a function of doping in its phase diagram. Here, doping, either intentional or accidental, usually takes place, for example, via cation substitution in $\mathrm{La_{2-x}Sr_xCuO_4}$ \cite{Wen2019}, or via inclusion of interstitial dopant oxygen atoms (Oi) in $\mathrm{Bi_2Sr_2CaCu_2O_{8+\delta}}$ \cite{McElroy2005}, $\mathrm{La_2CuO_{4+y}}$ \cite{Poccia2014} or $\mathrm{YBa_2Cu_3O_{6.5+y}}$ \cite{Ricci2014}. These can be treated as point-like scattering centers as well as extended defects that introduce disorder and deviate the neighboring atoms from their crystallographic positions. This poses a fundamental question regarding the context of the dome-shaped area of high temperature superconductivity in cuprates, on what mechanism is responsible for the reduction in $T_c$ upon overdoping as well as to the subsequent disappearance of superconductivity at a critical doping. Usually, this is ascribed to intrinsic effects, in which pairing correlations diminish with doping, due to screening of local Coulomb interactions \cite{Huang2017}, but some authours have also addressed the role of disorder in surpressing superconductivity \cite{Balatsky2006, Rullier2008, LeeHone2020}. Disorder, however, is usually incorporated as random on-site energies in Hubbard-like models that can lead to Anderson localization phenomena \cite{Peter2008, Nguyen2022, Nathan2021}, thus it is important to extend these effects to include also the possiblity of severe structural disorder within finite regions of the crystal.

One of the most significant results from the study of disorder effects in superconductivity is the well known Anderson's theorem, which states that both the transition temperature, $T_c$, and the isotropic gap, $\Delta_0$, of $s-$wave superconductors are insensitive to the presence of weak disorder at the mean-field level of BCS-like models \cite{Anderson1959, Abrikosov1958, Abrikosov1959}. One of the requirements of the theorem is that the density of states remains unchanged when compared to the pure metal case. If the influence of disorder is strong enough to deplete the density of states the theorem no longer holds and disorder dramatically affects superconductivity \cite{Cren2000}. {Furthermore, the effects of disorder in the superconducting state of cuprate superconductors is still a matter of debate. For instance, disorder was shown to enhance the mean-field superconducting temperature for systems with sign-changing order parameter, as in the case of cuprates \cite{Gastiasoro2018}. On the other hand, the pair field amplitude is shown to decrease with increased doping, in the vicinity of the doping-tuned quantum superconductor-to-metal transition, as in the case of the overdoped side of the phase diagram of cuprates \cite{Li2021}. On top of that, for $s-$wave superconductors with short coherence lenght, disorder can both enhance or suppress the critical temperature depending on electron doping \cite{Semenikhin2003}. Therefore it is clear that the treatment of disorder is important to characterize how the superconducting state is affected and we emphasize the significance of taking into account spatially extended defects, rather than only point-like impurity potentials.}

In the case of strong disorder and high concentration of impurity centers the superconducting correlation length is comparable to the disorder correlation length and the mean-field equations can lead to self-organized granularity where fluctuations of the local order parameter are present \cite{Dodaro2018}. This is likely to be the case for overdoped cuprate superconductors with high concentration of interstitial oxygens that can lead to the formation of nanosized oxygen puddles, regions with agglomeration of Oi, that support superconductivity \cite{Poccia2014, Ricci2014, Campi2013, Ricci2013, Poccia2020}, {but it is also relevant for other systems where granularity and the division between superconducting and metallic domains is important, for instance $\mathrm{Mg_{1-x}Al_xB_2}$ \cite{Conradson2009}, thin films of NbN doped with magnetic impurities \cite{Adhikari2022, Jha2013} and in disordered $\mathrm{InO_x}$ films \cite{Lewellyn2020}. Remarkably, disorder also induces reduction of $T_c$ in molybdenum nitride ($\mathrm{Mo_2N_x}$) thin films by increasing the amount of amorphous $\mathrm{MoN}$ regions \cite{Haberkorn2018} and the loss of superconudctivity is associated with the increase of dopant $\mathrm{Al}$ atoms near a structural instability in $\mathrm{Al}-$doped $\mathrm{MgB_2}$ \cite{Slusky2001}. Overall, we see that the understanding of the effects of spatially large disorder may be important to investigate not only their importance in the superconducting state of cuprates, but also in different classes of superconducting materials.} 

The case of {\it unconventional} high temperature $d-$wave superconductivity in hole-doped cuprates has been of experimental and theoretical significance since its discovery \cite{Muller1986}. Apart from several different physical characteristics, one of the main differences between these materials and the {\it conventional} BCS superconductors is that the superconducting gap amplitude is not homogenous. This is evidenced by scanning tunneling microscopy (STM) spectra in $\mathrm{Bi_2Sr_2CaCu_2O_{8+\delta}}$ at different doping levels, where the inhomogenous gap in the superconducting regime is revealed to be represented by a variety of gap sizes and amplitudes occuring in all samples as the concentration of dopants is varied \cite{McElroy2005}. Most remarkably, there is a clear correlation between the position of Oi agglomerates and the amplitudes of the gaps, since regions with larger groups of dopants are observed to correspond to regions of larger gap amplitudes \cite{McElroy2005}, {which is consistent with the picture of local superconductivity induced by disorder, that is also revealed by STM in other materials, such as boron-doped granular diamond \cite{Zhang2013}}. In parallel, Oi dopants have been observed to self-organize into nanosized regions, or {\it puddles}, as mentioned above, via $\mu$XRS in $\mathrm{HgBa_2CuO_{4+\delta}}$ \cite{Campi2015}, as well as in other cuprate compounds \cite{Poccia2010}. Remarkably, it has been observed that spatial variations in the self-organization of the nanosized Oi-rich puddles have a direct effect on superconductivity, through variations in the critical temperature \cite{Ricci2014-1}. Therefore, it is of paramount importance a deeper understanding, from a theoretical perspective, of the role of the oxygen puddles in the physics of hole-doped cuprates {and generally the effects of sptially extended disorder in the superconducting state of granular systems.}

In this work, we aim to investigate the effects of how extended structural disorder caused by the agglomeration of {dopant atoms} in puddles is responsible for the appearence of finite (nonzero) center-of-mass (CM) momentum Cooper pairs. This derives from the previously postulated mechanism for the emergence of {unconventional} high-temperature superconductivity that invokes the phase synchronization of the networks of the superconducting puddles \cite{Velasco2020}. We first extend the puddle model to derive analytical expressions showing how the superconducting gap, and thus the critical temperature, are affected by the presence of Cooper pairs with finite CM momentum and structural disorder. Then we show numerically that both $T_c$ and $\Delta_0$ decrease with increasing disorder, in a {general} physical mechanism for the diminution and loss of superconductivity resulting from the reduction of the available phase space for Cooper pairing due to the development of a nonzero, finite CM momentum Cooper pairs {in the landscape of spatially extended disorder centers.}

This paper is divided as following: in Sec. \ref{Sec: 2} we describe the effects of structural disorder {and how this can be ascribed to effects that} the agglomeration of dopants within each puddle causes to the system. In Sec. \ref{Sec: 3} we explain the puddle model, which is the base for the calculations presented in this work, and derive the effective interaction between electrons and the network of puddles, giving rise to a finite CM momentum pair state.  In Sec. \ref{Sec: 4} we derive the the self-consistent equation for the amplitude of the superconducting gap in terms of disorder and finite CM momentum Cooper pairs. Section \ref{Sec: 5} is devoted to the numerical calculations. We conclude with a discussion of the implications of our results within the framework of networks of nano-sized puddles and summarize our findings in Sec. \ref{Sec: 6}.

\section{structural disorder}
\label{Sec: 2}

\begin{figure}[!t]
\includegraphics[width = \linewidth]{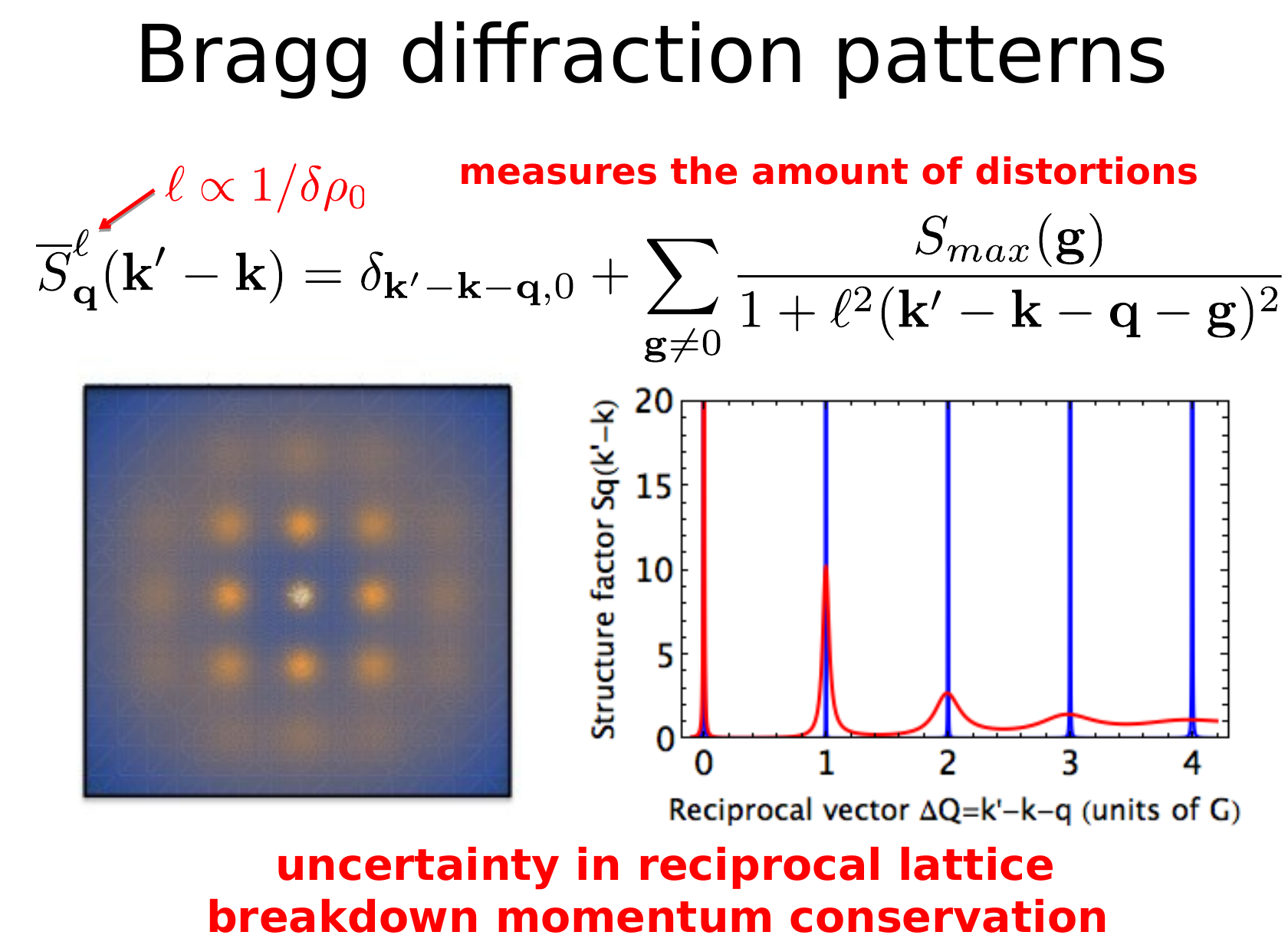}
\caption{{\it Top}: Structure factor for disordered media from Hosemann's paracrystalline theory \cite{Hosemann1950}, given by eq. (\ref{Disordered-Structure-Factor}) in the text. The pristine case corresponds to the $\ell\rightarrow\infty$ limit, where the structure factor is given by delta-peaks at reciprocal lattice vectors and momentum is conserved (here $\ell$ is a measure of disorder and for this reason should be inversely related to the residual resistivity shift due to structural disorder, $\ell\propto 1/\delta\rho_0$). {\it Bottom left:}: $-$ Bragg diffraction pattern for a structually disordered medium, showing Bragg peaks at the central region and Bragg rings at the outter region; {\it Bottom right} $-$ plot of the structure factor as a function of momentum transfer, $\Delta{\bf Q}$, showing well defined Bragg peaks, for small momentum transfer, at the reciprocal lattice vectors, $\mathbf{G}$, while the Bragg peaks become ever broader, at larger momentum transfer, eventually merging into rings.}
\label{Fig: Structure-Factor}
\end{figure}

{Before we proceed to the mathematical discussion regarding the rising of a superconducting state with Cooper pairs showing finite CM momentum, it is important to briefly discuss which kind of disorder is physically inducing it}. In order to do that, we introduce concepts arising from the study of structural disorder, which is the kind of perturbation that the agglomeration of Oi causes in the crystalline structure of different cuprate systems, as for example by tilting the $\mathrm{CuO_6}$ octahedra in $\mathrm{La_2CuO_{4+\delta}}$ \cite{Zhang2022} and by altering the distance between the apical oxygen and the planar copper atom in $\mathrm{Bi_2Sr_2CaCu_2O_{8+\delta}}$ \cite{Slezak2008}, {but it also has effects in the superconducting state, since it is the structural disorder that is responsible for the reduction of $T_c$ in $\mathrm{Nd_{1+x}Ba_{2-x}Cu_3O_z}$ $(z \approx 7)$ \cite{Petrykin2000} and in a different superconducting system, as for instance $\mathrm{Mo_2N_x}$ thin films \cite{Haberkorn2018}.}

{Translational invariance is one of the most fundamental properties of pristine crystals. A perfect crystal is characterized by very intense and sharp peaks in the Fraunhofer diffraction pattern of Bragg scattering experiments. The existence of such sharp peaks follows directly from Heisenberg's uncertainty principle and their location is determined by the crystalline-lattice structure factor. For a pristine crystal all atoms are at their ideal locations and the structure factor is characterized by $S({\bf k}^\prime-{\bf k})=\sum_{{\bf g}}\delta_{{\bf k}^\prime-{\bf k},{\bf g}}$, where ${\bf g}$ is a reciprocal lattice vector. The Fraunhoffer diffraction pattern in this case corresponds to $\delta-$like peaks as shown in Fig. \ref{Fig: Structure-Factor} and the kinematic constraint of quasi-momentum conservation, ${\bf k}^\prime={\bf k}+{\bf g}$,
forms the basis for Bloch's theorem. In the opposite limit of a random atom gas, however, an extended Bloch wave with well defined momentum state, ${\bf k}$, that interacts with ions located at a particular, well defined position of the crystal (zero uncertainty $\Delta{\bf r}\rightarrow 0$), scatters into another extended Bloch wave with momentum state, ${\bf k}^\prime$, with infinite uncertainty, $\Delta{\bf k}\rightarrow\infty$. In this case the structure factor takes the form $S({\bf q})=1$. There are no kinematic constraints whatsoever relating ${\bf k}$ and ${\bf k}^\prime$ to ${\bf g}$ and the Fraunhoffer diffraction pattern in this case corresponds to an isotropic disc of even intensity, as shown in Fig. \ref{Fig: Structure-Factor}.
}

{Interpolating between the pristine and random limits described above by increasing disorder is pivotal to the description of inherently inhomogeneous systems. If disorder is of the first type, namely weak disorder, all atoms deviate only slightly from their ideal positions in the crystal, independently of the deviations of their neighbors \cite{Dullens2007}. This is the case of pointlike defects, thermal vibrations or micro-mechanical strains, and this kind of disorder preserves long range crystalline order. In this case the widths of the peaks in the Fraunhoffer diffraction pattern are not affected. If disorder is of the second type, namely strong disorder, however, the atoms deviate significantly from their ideal positions in the crystal, and deviations amongst neighboring atoms are correlated. This is the case of extended defects, amorphous regions, molten materials, etc, and this type of disorder causes the loss of long range crystalline order. In these paracrystalline structures, not only the intensity of the diffraction peaks will decrease but, most importantly, their widths will suffer from a nonlinear increase of their integral breadth, $\delta {\bf g}$, for successive orders of Bragg reflections. The complete paracrystalline theory was proposed by Hosemann \cite{Hosemann1950}. Hosemann included fluctuations that introduce correlations between pairs of atoms and decrease with separation, ultimately causing the peaks in the structure factor of the material to broaden the larger the reciprocal lattice. The result is a structure factor composed by a sum of Lorentzians \cite{Hosemann1995}}

\begin{equation}
\overline{S}_{\bf q}({\bf k}^\prime-{\bf k})=\sum_{{\bf g}}\frac{S_{max}({\bf g})}{1+\ell^2_{hkl}({\bf q}-{\bf k}^\prime+{\bf k}-{\bf g})^2},
\label{Disordered-Structure-Factor}
\end{equation}
of amplitudes $S_{max}({\bf g})=4/\sigma^2{\bf g}^2$ and breadths for Bragg reflections, $|\delta {\bf g}|\equiv 1/\ell_{hkl}=\sigma^2\pi^2(h^2+k^2+l^2)/a_0$, given in terms of the original lattice parameter $a_0$ and the momentum transfer, ${\bf q}$. Hosemann's paracrystalline theory allows us then to interpolate continuously between pristine and random cases through the fluctuation parameter $\sigma$:
\begin{itemize}
    \item for $\sigma\rightarrow 0$ we have $\ell_{hkl}\rightarrow \infty,\forall h,k,l$ and we obtain $\overline{S}_{\bf q}({\bf k}^\prime-{\bf k})=\sum_{{\bf g}}\delta_{{\bf q},{{\bf k}^\prime-{\bf k}}+{\bf g}}$, enforcing the kinematic constraint of momentum conservation, ${\bf q}={\bf k}^\prime-{\bf k}+{\bf g}$, typical of pristine crystals \cite{Hosemann1995};
    \item for $\sigma\rightarrow \infty$ we have $\ell_{hkl}\rightarrow 0,\forall h,k,l$ and we end up with $\overline{S}_{\bf q}({\bf k}^\prime-{\bf k})=S_{max}({\bf 0})\rightarrow 1$, isotropic, for arbitrary ${\bf q}, {\bf k}, {\bf k}^\prime$ and determined solely by the ${\bf g}=0$ contribution, typical of infinite, aperiodic systems \cite{Hosemann1995};
    \item for $0\leq\sigma\leq\infty$ we have $\infty\geq\ell_{hkl}\geq 0$ and the structure factor, $\overline{S}_{\bf q}({\bf k}^\prime-{\bf k})$, will be composed by sharp Bragg peaks at small ${\bf g}$ (large $\ell_{hkl}$) and isotropic discs for larger ${\bf g}$ (small $\ell_{hkl}$), as shown in Fig. \ref{Fig: Structure-Factor}, relaxing the kinematic constraint of momentum conservation, ${\bf q}\not\approx{\bf k}^\prime-{\bf k}+{\bf g}$, typical of a paracrystal, liquids, strongly disordered or amorphous systems \cite{Hosemann1995}.
\end{itemize}
{Following the characterization of the effects of disorder in the momentum conservation, Bergmann \cite{Bergmann1971} introduced the idea that disorder of the 2nd kind would provide Cooper pairs with a finite center-of-mass momentum, through a process coined as {\it pseudo-Umklapp scattering}, supressing the available phase space for pairing and thus reducing the size of the gap. The thorough and complete mathematical description of this process was recently revisited by Ref. \cite{MBSN2021}, where a whole large class of both conventional and unconventional superconducting materials were considered. Physically, finite center-of-mass momentum Cooper pairs arise from the relaxation of the kinematic constraint for momentum conservation in the case of strongly inhomogeneous systems, in other words, the case of second-type disorder. Accordingly, as we shall see below, the effects of treating disorder as extended defects has an important impact in the superconducting state of systems that can be described as an intercalation between superconducting islands and a metallic environment, where the islands are rich in dopant atoms.}

\section{Inhomogeneous dopant puddles}
\label{Sec: 3}

\begin{figure}
\includegraphics[width = \linewidth]{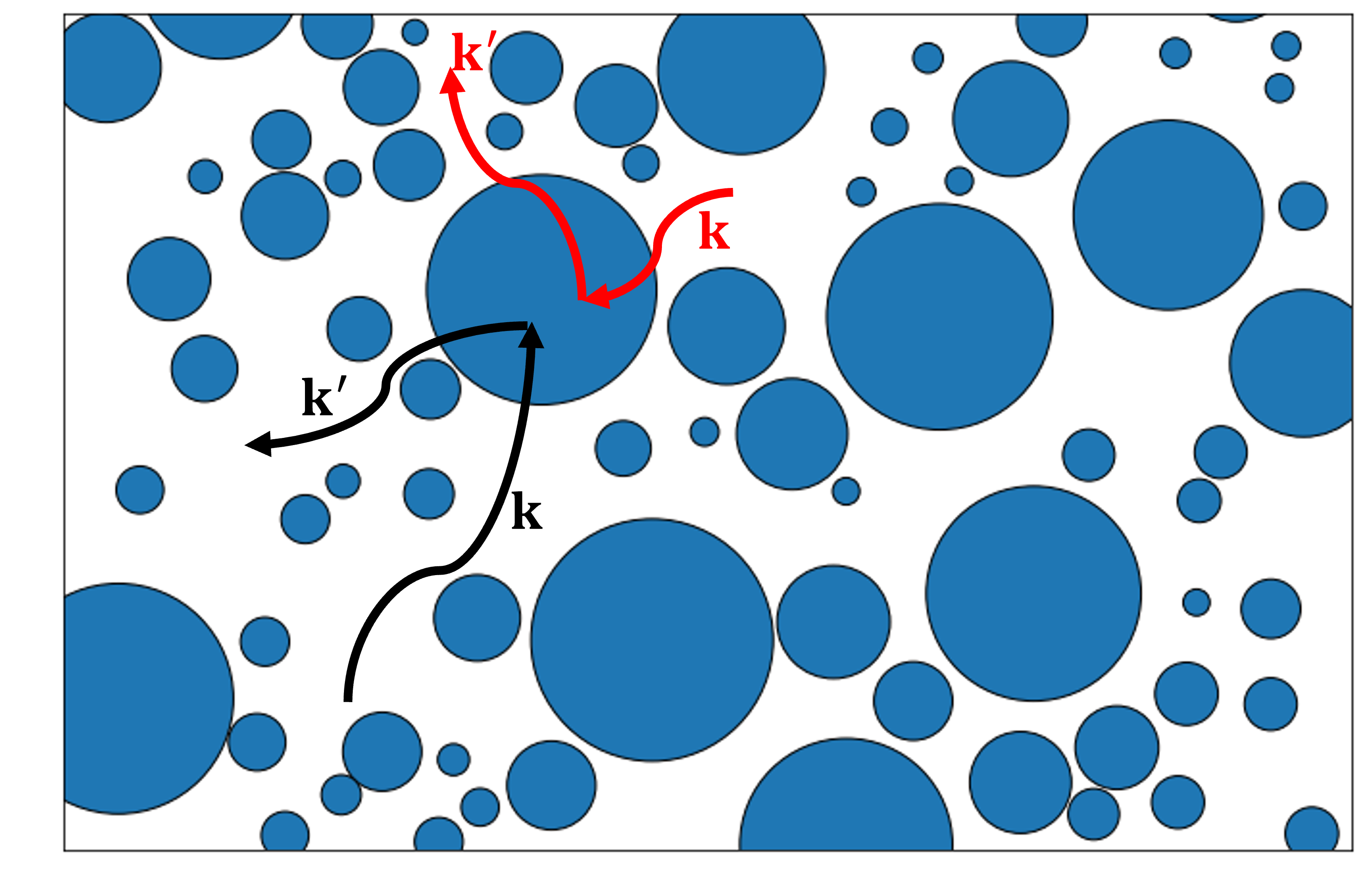}
\caption{Pictorical view of the disordered background introduced by the network of puddles (blue) in the system. The network consists of puddles of different sizes, defined by the radius of each insertion. Electrons (black and red) scatter in each puddle and, in the superconducting state, percolate within the network.}
\label{Fig: Puddles}
\end{figure}

The oxygen rich nanopuddles have different elastic properties than their surroundings, and can therefore be considered as elastic insertions in an otherwise homogeneous medium, with its own vibrational mode, forming a network of superconducting nanoscale puddles, as shown in Fig. \ref{Fig: Puddles}. In terms of the Kuramoto model for sychronization of phase oscillators \cite{Velasco2020, Kuramoto1975, Kuramoto1987}, each nanosized puddle is assigned to a phase, that in the underdoped regime evolves independently of the others, giving rise to localized patches of superconductivity, as revaled by STM and other techniques. With increased concentration of, for example Oi as in the case of cuprates, through doping, the superfluid density is responsible for the enhancement of the interactions between the puddles and, in terms of the Kuramoto model, to lock their phases in a synchronous way. Following a BCS-like procedure, the order parameter for synchronization is connected to the amplitude of the bulk superconductor gap, that is non zero only after the locking of the global phase in the synchronized phase. The synchronization and the large frequency of the global network of puddles is also responsible for large values of $T_c$ in the optimally doped cuprates \cite{Velasco2020}.

Inspired by these experimental and theoretical findings, we introduce a model Hamiltonian that captures the interaction between electrons and localized vibrations that arise from the agglomeration of dopant atoms in one puddle. This interaction must be local, since each electron will only interact with the quantized vibration whenever it is in the region defined by the puddle (see Fig. \ref{Fig: Puddles}). The minimal model that captures this physical situation can be divided in $H = H_{el} + H_{p} + H_{el-p}$, with

$$H_{el} = \sum_{\mathbf{k},\sigma} \xi_{\mathbf{k}} c^{\dagger}_{\mathbf{k}, \sigma} c_{\mathbf{k}, \sigma} + \sum_{\mathbf{k,k^\prime}}\mathcal{T}_{\mathbf{k,k^\prime}}c^{\dagger}_{\mathbf{k^\prime},\sigma}c_{\mathbf{k},\sigma},$$
where the first term represents a band of electrons with dispersion $\xi_{\mathbf{k}}$ measured relative to the chemical potential, with creation $c^{\dagger}_{\mathbf{k},\sigma}$ and annihilation $c_{\mathbf{k}, \sigma}$ fermionic operators. The second term represents the scattering of electrons in each inhomogeneity, with strenght controled by the spin-preserving momentum transfer disorder matrix $\mathcal{T}_{\mathbf{k,k^\prime}}$. The {dopant puddles} are described by local phonon modes 

$$H_{p} = \sum_{\mathbf{q}}\hbar\omega_{\mathbf{q}}a^{\dagger}_{\mathbf{q}}a_{\mathbf{q}},$$
with frequencies $\omega_{\mathbf{q}}$ and the creation ($a^\dagger_{\mathbf{q}}$) and annihilation ($a_{\mathbf{q}}$) bosonic operators, responsible for the description of the localized vibration of each puddle. Finally, the interaction term can be described as

$$H_{el-p} = \sum_{\mathbf{r, R},\sigma} g(\mathbf{r - R})c^{\dagger}_{\mathbf{r}, \sigma}c_{\mathbf{r},\sigma}\left(a_{\mathbf{R}}^{\dagger} + a_{\mathbf{R}}\right), $$
where $\mathbf{r}$ and $\mathbf{R}$ are the electron and puddle locations, respectively. The puddle is a finite size region in space, thus $\mathbf{R}$ defines the center of this region that can be modeled as a sphere. The interaction strenght $g(\mathbf{r-R})$ is only relevant whenever the electron is in the region around the puddle, which can be modeled using a Gogny-type short range interaction that is dependent on the radius of the dopant agglomeration region \cite{Gogny1975}. After performing the transformation to momentum space, the interaction term is written as

\begin{equation}
    H_{el-p} = \sum_{\mathbf{k,k^\prime,\sigma,q}} M(\mathbf{q,k-k^\prime})c^{\dagger}_{\mathbf{k},\sigma}c_{\mathbf{k^\prime},\sigma}\left(a^\dagger_{\mathbf{-q}} + a_{\mathbf{q}}  \right),
    \label{Eq: 1}
\end{equation}
where $$M(\mathbf{q,k-k^\prime}) = \sum_{\mathbf{R}} g(\mathbf{k-k^\prime})\exp{[i(\mathbf{q - [k - k^\prime]})\cdot \mathbf{R}]}$$ is associated with the fact that the puddles are not present in all sites, rather they are inhomogeneously distributed around the system, thus the summation has to be retained only to these regions, which is relevant for the case of $\mathrm{Bi_2Sr_2CaCu_2O_{8+\delta}}$, since locations of dopant oxygens are observed to be consistent with the position inferred from local strain analysis of the incommensurate structure, as imaged by scanning transmission electron microscopy (STEM) \cite{Song2019}, which means that the crucial oxygen dopants are periodically distributed in correlation with local strain. However, not all strained regions are occupied with dopant oxygen atoms, that is the distribution of Oi is inhomogeneous, which justifies our approximation and is consistent with STM measurements \cite{Zeljkovic2014, Zhang2013}. In the limits of a clean or a totally doped system, this term can be treated exactly. The factor $g(\mathbf{k-k^\prime})$ is the Fourier transform of the interacting potential between the electrons and the puddles and controls the momentum transfer between the incoming and scattered electron.

One can see from Eq. \eqref{Eq: 1} that the presence of a finite density of puddles spread around the system give rise to an off-diagonal term associated with the momentum transfer $\mathbf{k-k^\prime}$ that comes from the interacting potential. In the limit that the summation over $M(\mathbf{q,k-k^\prime})$ can be made exactly, one recovers the usual definition of an electron-phonon interaction, where the momentum transfer is the momentum of the local phononic mode $\mathbf{q}$, as in the Frohlich \cite{Frohlich1937} and Holstein \cite{ Holstein1959} models, for example. In order to explore the effects of this kind of interaction in the form of pairing, we introduce an unitary transformation $H^\prime = e^{-S}He^{S}$, with an {\it ansatz} for the transformation matrix

\begin{equation}
    S = \sum_{\mathbf{k,k^\prime,\sigma,q,Q}} M(\mathbf{q,k-k^\prime})c^{\dagger}_{\mathbf{k},\sigma}c_{\mathbf{k^\prime},\sigma}\left(xa^\dagger_{\mathbf{-q}} + ya_{\mathbf{q}}  \right),
    \label{Eq: S}
\end{equation}
where $x$ and $y$ are factors determined {\it a posteriori}. After the transformation (see Appendix \ref{App: A} for details), we end with an effective interaction written as

\begin{equation}
    H_{\mathrm{eff}} = \sum_{\mathbf{k}, \mathbf{k}^{\prime}} \sum_{\mathbf{p}, \mathbf{p}^{\prime}} V(\mathbf{k,k^\prime})f(\mathbf{p,p^\prime})c_{\mathbf{k}, \uparrow}^{\dagger} c_{\mathbf{p}-\mathbf{k}, \downarrow}^{\dagger} c_{\mathbf{p}^{\prime}-\mathbf{k}^{\prime}, \downarrow} c_{\mathbf{k}^{\prime}, \uparrow}
    \label{Eq: 2}
\end{equation}
with $V(\mathbf{k,k^\prime}) = D(\mathbf{k,k^\prime})|g(\mathbf{k - k^\prime})|^2$ being the potential arising from the interaction between electrons and the {spacially extended inhomogeneities}, $D(\mathbf{k,k^\prime})$ the phononic propagator associated with the local phonon modes produced by the vibrating puddles and $f(\mathbf{p,p^\prime}) = \sum_{\mathbf{R}} e^{-i\left(\mathbf{p-p^\prime}\right)\cdot \mathbf{R}}$ the phase factor controlling momentum transfer between the interacting electrons. {A finite center-of-mass momentum is generated due to the broadening of the momentum transfer control function $f(\mathbf{p,p^\prime})$. While for clean systems $f(\mathbf{p,p^\prime})$ is a $\delta-$function enforcing momentum conservation, $\mathbf{p=p^\prime}$, for inhomogeneous media $f(\mathbf{p,p^\prime})$ becomes broadened, relaxing the kinematic constraints and allowing $\mathbf{p\neq p^\prime}$.} In the regime where the phononic propagator is negative, given that $\xi_\mathbf{k} \approx \xi_\mathbf{k^\prime}$, we have an effective attractive interaction between the electrons mediated by the nanopuddles. Remarkably, this interaction leads to the formation of finite center-of-mass momentum Cooper pairs represented by $\mathbf{p}$ and $\mathbf{p^\prime}$. Therefore, from the perspective of inhomogeneously distributed puddles bringing {spacially extended} disorder to an otherwise clean medium, a bound state between two electrons can be formed with a finite center-of-mass momentum that is associated with the strenght of the interaction between the electrons forming the pair and the agglomeration of dopant atoms in one nanopuddle.

It is important to notice that the states arising from the effective Hamiltonian in Eq. \eqref{Eq: 2} are different from other proposed pair states with finite CM momentum, as for example the Fulde-Ferrell-Larkin-Ovchinnikov (FFLO) state, where finite center-of-mass momentum Cooper pairs can be stabilized under a finite magnetic field via the Zeeman coupling \cite{Fulde1964, Larkin1965}, and the recently proposed current driven FFLO state \cite{Doh2006}. Moreover, it has been shown that, even without the presence of a magnetic field or other external potentials, a finite CM momentum Cooper pair can be stable in a superconducting ground state as pointed in Ref. \cite{Loder2010}, but the authors do not explore the effects that can give rise to this kind of state. Here we start from the fact that nanosized puddles are formed via doping and the responsible for the CM momentum of the pairs is {the extension in space of the} disorder induced by the puddles in the system.

Eventhough we are not considering any specific form for the interaction potential $g(\mathbf{k-k^\prime})$, it is important to comment that the only requirement is that it must be a finite size potential in real space, which means that is not a point-like disorder center that is scattering the electrons in the interaction term of Eq. \eqref{Eq: 1}, rather is a region in space defined by the agglomeration of oxygen interstitials, {in the specific case of cuprates, or any other dopant atoms entering the homogeneous media}. In this case, we can point to potentials like the Woods-Saxon potential \cite{Woods1954} that is used to describe the forces applied on protons and neutrons in the atomic nucleous or the Gogny-type interactions \cite{Gogny1973, Gogny1975_1, Decharge1980}, which is another kind of nucleon-nucleon potential that has also found applications in astrophysics \cite{Gonzalez2018}, as possible candidates to describe the electron-puddle interaction. However, a precise and detailed description of such potential would required more knowledege about the formation of the nanosized puddles and its effects on the crystal structure of the host material, which would affect the electronic degrees of freedom \cite{He2006}, but this is outside the scope of the present study.

\section{Disorder and gap fluctuations}
\label{Sec: 4}
We now address how the superconducting state of the effective interaction derived in Sec. \ref{Sec: 2} is affected by the structural disorder effects introduced in the previous section. We start from the effective Hamiltonian in Eq. \eqref{Eq: 2} and, within a mean-field decoupling of the quartic term, write the equation for the superconducting gap as

\begin{equation}
\Delta_{\mathbf{k}}=-\sum_{\mathbf{k}^{\prime}, \mathbf{p}^{\prime}} V_{\mathbf{k}, \mathbf{k}^{\prime}} f_{\mathbf{0}, \mathbf{p}^{\prime}}\left\langle c_{\mathbf{p}^{\prime}-\mathbf{k}^{\prime} \downarrow} c_{\mathbf{k}^{\prime} \uparrow}\right\rangle,
\label{Eq: 3}
\end{equation}
where we set $\mathbf{p} = 0$, since we want to describe amplitude fluctuations for the superconducting gap in the Cooper channel. For the superconducting state formed by singlet pairs with finite CM momentum, the system can be represented by the spin-independent imaginary time Green's function $\mathcal{G}(\mathbf{k,k^\prime,\tau}) = -\left\langle T_{\tau} c_{\mathbf{k},\sigma}(\tau)c^\dagger_{\mathbf{k^\prime},\sigma}(0) \right\rangle$ and the anomolous pair propagators $\mathcal{F}(\mathbf{k,k^\prime,\tau}) = \left\langle T_{\tau}c_{\mathbf{k},\sigma}(\tau)c_{\mathbf{k^\prime}\sigma^\prime}(0) \right\rangle$ and $\mathcal{F}^*(\mathbf{k,k^\prime},\tau) = \left\langle T_{\tau}c^\dagger_{\mathbf{k},\sigma}(\tau) c^\dagger_{\mathbf{k^\prime},\sigma^\prime}(0) \right\rangle$ for $\sigma \neq \sigma^\prime$. Within Nambu's formalism, we can write the decoupled effective Hamiltonian from Eq. \eqref{Eq: 2} and the electronic components from $H_{el}$ in matrix form and derive in first order perturbation theory the electronic Green's function for an inhomogeneous system with {spatially extended} disorder as

\begin{eqnarray}
\mathbf{G}\left(\mathbf{k}, \mathbf{k}^{\prime}, i \omega_n\right) & = & \mathbf{G}_0\left(\mathbf{k}, \mathbf{k}^{\prime}, i \omega_n\right)\nonumber\\
&+&\sum_{\mathbf{p}, \mathbf{p}^{\prime}} \mathbf{G}_0\left(\mathbf{k}, \mathbf{p}, i \omega_n\right) \mathcal{T}_{\mathbf{p}, \mathbf{p}^{\prime}} \sigma_3 \mathbf{G}\left(\mathbf{p}^{\prime}, \mathbf{k}^{\prime}, i \omega_n\right),\nonumber
\end{eqnarray}
where $\mathbf{G}_0\left(\mathbf{k}, \mathbf{k}^{\prime}, i \omega_n\right)$ is the matrix form of the translationally invariant electronic Green's function in frequency space, $i\omega_n$ are the fermionic Matsubara frequencies and $\sigma_3$ is a Pauli matrix. The diagonal elements of this matrix are defined by the bare Green's function in the superconducting state, $\mathcal{G}_0(\mathbf{k},i\omega_n)$, and its off-diagonal terms are represented by the anomalous propagators $\mathcal{F}_0(\mathbf{k},i\omega_n)$ which are written as
$$
\begin{aligned}
&\mathcal{G}_0\left(\mathbf{k}, i \omega_n\right)=\frac{-\left(i \omega_n+\xi_{\mathbf{k}}\right)}{\omega_n^2+\xi_{\mathbf{k}}^2+\left|\Delta_{\mathbf{k}}\right|^2}, \\
&\mathcal{F}_0\left(\mathbf{k}, i \omega_n\right)=\frac{\Delta_{\mathbf{k}}}{\omega_n^2+\xi_{\mathbf{k}}^2+\left|\Delta_{\mathbf{k}}\right|^2}.
\end{aligned}
$$
In order to proceed, we shall take a couple of approximations: first we consider the case of {an overdoped sample}, which puts the system in a high concentration of disorder, thus $\mathcal{T}_{\mathbf{p,p^\prime}} = \mathcal{T}f(\mathbf{p,p^\prime})$, where disorder influences the momentum transfer controled by the phase factor $f(\mathbf{p,p^\prime})$ with strenght $\mathcal{T}$. Second we assume that for a translationally invariant system the normal and anomalous Green's functions can be rewritten as $\mathcal{G}_0(\mathbf{k,k^\prime,i\omega_n}) = \mathcal{G}_0(\mathbf{k},i\omega_n)\delta_{\mathbf{k,k^\prime}}$ and $\mathcal{F}_0(\mathbf{k,k^\prime,i\omega_n}) = \mathcal{F}_0(\mathbf{k},i\omega_n)\delta_{\mathbf{-k,k^\prime}}$. Following these couple of approximations, the first order pertubation theory expansion of the interacting Green's function is simplified 

\begin{eqnarray}
 \mathbf{G}\left(\mathbf{k}, \mathbf{k}^{\prime}, i \omega_n\right) &=&\mathbf{G}_0\left(\mathbf{k}, i \omega_n\right) \delta_{\mathbf{k}, \mathbf{k}^{\prime}}\nonumber\\
 &+&\mathcal{T} f_{\mathbf{k}, \mathbf{k}^{\prime}} \mathbf{G}_0\left(\mathbf{k}, i \omega_n\right) \sigma_3 \mathbf{G}_0\left(\mathbf{k}^{\prime}, i \omega_n\right).
 \label{Eq: 4}
\end{eqnarray}
From the gap equation in Eq. (\ref{Eq: 3}) and from the definition of the anomalous propagator, we write

\begin{eqnarray}
    \Delta_{\mathbf{k}} &=&-\sum_{\mathbf{k}^{\prime}, \mathbf{p}^{\prime}} V_{\mathbf{k}, \mathbf{k}^{\prime}} f_{\mathbf{0}, \mathbf{p}^{\prime}}\left\langle c_{\mathbf{p}^{\prime}-\mathbf{k}^{\prime} \downarrow} c_{\mathbf{k}^{\prime} \uparrow}\right\rangle\nonumber\\
    &=&-\sum_{\mathbf{k}^{\prime}, \mathbf{p}^{\prime}} V_{\mathbf{k}, \mathbf{k}^{\prime}} f_{\mathbf{0}, \mathbf{p}^{\prime}}\left[\frac{1}{\beta} \sum_{\omega_n} \mathcal{F}\left(\mathbf{p}^{\prime}-\mathbf{k}^{\prime}, \mathbf{k}^{\prime}, i \omega_n\right)\right],
    \label{Eq: 5}
\end{eqnarray}
with $\beta = 1/T$ being the inverse temperature (in units of $k_B = 1$). By using the matrix form in Eq. \eqref{Eq: 4}, we get the form of the interacting anomalous propagator, where it is worth noting that the normal and anomalous propagators mix in the impurity scattering. Despite the anomalous Green's function being invariant for time reversal, the normal one is not, and since disorder produces the transformation $\mathcal{F}_0(\mathbf{k},i\omega_n) \leftrightarrow \mathcal{G}_0(\mathbf{k},i\omega_n)$ we clearly see this is a mechanism that breaks time reversal invariance. As a consequence, this mechanism breaks the Cooper pair that leaks into the normal metal surrounding the puddles.

In order to understand the effects of disorder and finite CM momentum in the gap equation, we substitute the form of the anomalous propagator given by the matrix in Eq. \eqref{Eq: 4} inside Eq. \eqref{Eq: 5} to write the gap equation as $\Delta_{\mathbf{k}} = \Delta_{\mathbf{k}}^{\mathrm{BCS}} + \delta\Delta_{\mathbf{k}}$, where 

\begin{eqnarray}
\Delta_{\mathbf{k}}^{\mathrm{BCS}} & = & -\sum_{\mathbf{k^\prime}}\frac{V_{\mathbf{k,k^\prime}}\Delta_{\mathbf{k^\prime}}}{2E_{\mathbf{k^\prime}}}\tanh{\left(\frac{\beta E_{\mathbf{k^\prime}}}{2}\right)},
\end{eqnarray}
is the BCS limit for the gap equation, arising from the first term in Eq. {\eqref{Eq: 4}}, with the bare anomolous propagators and $E_{\mathbf{k}} = \sqrt{\xi^2_{\mathbf{k}} +\Delta^2_{\mathbf{k}}}$. Then

\begin{eqnarray}
    \delta\Delta_{\mathbf{k}}  = \mathcal{T} \sum_{\mathbf{k}^{\prime}, \mathbf{p}^{\prime}} V_{\mathbf{k}, \mathbf{k}^{\prime}}f_{\mathbf{0}, \mathbf{p}^{\prime}} f_{\mathbf{p}^{\prime}, \mathbf{0}}\frac{1}{\beta} \sum_{\omega_n}\left\{\mathcal{F}_0\mathcal{G}_0+\mathcal{G}_0\mathcal{F}_0\right\}
    \label{Eq: 8}
\end{eqnarray}
is the correction to the superconductor gap due to effects of disorder in the system. {We point that even though the splitting of the gap equation in two factors seems to induce the introduction of two different order parameters, this division comes solely from the fact that the anomalous propagator has a bare contribution and a factor proportional to disorder in the perturbation theory. As we show in the next section, during the numerical calculations we introduce only one order parameter that characterizes the superconducting state.} The factor $\left[f_{0,\mathbf{p^\prime}}f_{\mathbf{p^{\prime}},0}\right]$ can be treated within a mean over disorder in order to calculate the interference factor as $\overline{\left[f_{0,\mathbf{p^\prime}}f_{\mathbf{p^{\prime}},0}\right]} = \overline{|f_{0,\mathbf{p^\prime}}|^2}\rightarrow S(\mathbf{p}^\prime)$, where $S(0,\mathbf{p}^\prime)$ is the static structure factor. Thus, the correction to the gap equation can be written in terms of the structure factor and we see that fluctuations associated with small CM momentum $\mathbf{p}^\prime \rightarrow 0$ are absent, since the structure factor $S(\mathbf{p^\prime}) \rightarrow 0$ and the gap equation is dominated by the BCS contribution. On the other hand, fluctuations associated with a finite center-of-mass momentum dominate over the BCS contribution when $\mathbf{p^\prime} \gg 0$ and $S(\mathbf{p^\prime}) \rightarrow 1$. In a general manner, the structure factor can be written as a sum of Lorentzians with peaks in wave vectors of the reciprocal lattice, as discussed in Sec. \ref{Sec: 2} and shown in Fig. \ref{Fig: Structure-Factor}.

Finally, we proceed by taking the Matsubara summations over the set of mixed Green's functions as in Eq. {\eqref{Eq: 8}} to arrive at the correction in terms of the disorder strenght $\mathcal{T}$ and the finite CM momentum of the Cooper pairs $\mathbf{p^\prime}$ as

\begin{eqnarray}
\begin{aligned}
\delta\Delta_{\mathbf{k}} &= \mathcal{T} \sum_{\mathbf{k}^{\prime}, \mathbf{p}^{\prime}} V_{\mathbf{k}, \mathbf{k}^{\prime}} S\left(\mathbf{p}^{\prime}\right) \frac{1}{2}\left[\frac{\Delta_{\mathbf{k}^{\prime}, \mathbf{p}^{\prime}}}{E_{\mathbf{k}^{\prime}-\mathbf{p}^{\prime}}} \frac{\xi_{\mathbf{k^\prime}}}{E_{\mathbf{k^\prime}}}+\frac{\Delta_{\mathbf{k}^{\prime}}}{E_{\mathbf{k}^{\prime}}} \frac{\xi_{\mathbf{k}^{\prime}-\mathbf{p}^{\prime}}}{E_{\mathbf{k}^{\prime}-\mathbf{p}^{\prime}}}\right] \\
& \times\left\{\frac{E_{\mathbf{k}^{\prime}-\mathbf{p}^{\prime}} \tanh \left(\frac{\beta E_{\mathbf{k}^{\prime}}}{2}\right)-E_{\mathbf{k}^{\prime}} \tanh \left(\frac{\beta E_{\mathbf{k}^{\prime}-\mathbf{p}^{\prime}}}{2}\right)}{E_{\mathbf{k}^{\prime}-\mathbf{p}^{\prime}}^2-E_{\mathbf{k}^{\prime}}^2}\right\}.
\end{aligned}
\label{Eq: 9}
\end{eqnarray}
%

It is importance to notice the dependence of the correction on the structure factor $S(\mathbf{p^\prime})$ controlling momentum transfer. In the limit of small amount of disorder, the so called first-type disorder \cite{Dullens2007}, as discussed in Sec. \ref{Sec: 2}, pointlike defects does not affect the BCS gap, in accordance with Anderson's Theorem, as we shall see in the next section. On the other hand, in the limit of high concentration of puddles, the system is in the limit of second-type disorder, associated with strain-induced lattice deformations, and both the amplitude of the superconducting gap and the critical temperature are affected. 

In order to proceed to the numerical analsysis, we perform an approximation for the structure factor based on the limits of disorder discussed above. For the first-type disorder, we choose $S(\mathbf{p^\prime}) = \delta_{0,\mathbf{p^\prime}}$, since no momentum transfer will be associated with pairs with finite CM momentum in the dilute limit. On the other hand, for the second-type disorder, we write $S(\mathbf{p^\prime}) = 1$, assuming a system with high concentration of {spatially extended disordered centers, forming the network of puddles}. These two limits for the disorder of the 1st and 2nd types can be understood as a {\it hard cutoff} for the CM momentum distribution within the structure factor and are made to simplify Eq. \eqref{Eq: 9} to the following numerical analysis.

\section{Numerical analysis}
\label{Sec: 5}
In order to fully understand the effects of disorder and CM momentum of the Cooper pairs in the superconducting gap amplitude we perfom a numerical integration of Eq. {\eqref{Eq: 9}}.  We use the decomposition $V_{\mathbf{k,k^\prime}} = -V_0\eta(\mathbf{k})\eta(\mathbf{k^\prime})$ and $\Delta_{\mathbf{k}} = \Delta_0\eta(\mathbf{k})$, where $\eta(\mathbf{k}) = \cos k_x - \cos k_y$ is a $d-$wave form factor, which gives the amplitude fluctuations of the order parameter with the same symmetry. When stated for comparison, we shall also use $V_{\mathbf{k,k^\prime}} = -V_0$ and $\Delta_{\mathbf{k}} = \Delta_0$ when considering a $s-$wave symmetry for the interaction and the gap. For the calculations in the square lattice, we consider a two-dimensional electronic dispersion with nearest- and next-nearest-neighbor hopping elements $(t,t^\prime)$ as

\begin{figure}[!t]
\includegraphics[width = \linewidth]{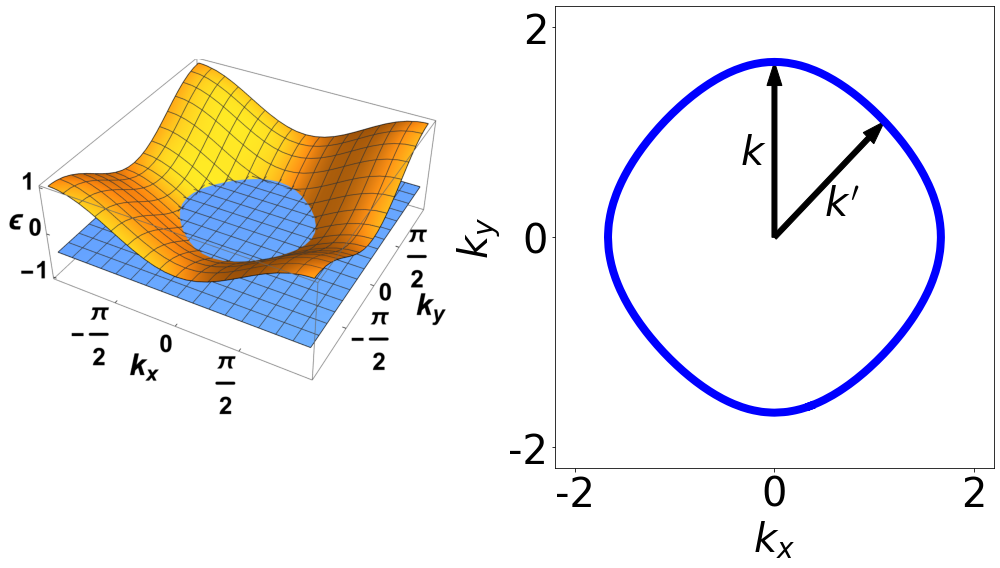}
\caption{Fermi surface structure used in calculations. {\it Left:} The 3D plot of Eq. {\eqref{Eq: 10}} in the first Brillouin zone in yellow and the chemical potential cut defining the Fermi level in blue. {\it Right:} The Fermi level defined by the cut at $\mu/4t = -0.45$. The vectors $\mathbf{k}$, fixed in the direction $(0, \pi)$, and $\mathbf{k^\prime}$, varying across the Fermi surface, are also shown.}
\label{Fig: Fermi_Surface}
\end{figure}

\begin{eqnarray}
    \epsilon_{\mathbf{k}} = -2t\left(\cos k_x + \cos k_y\right) + 4t^\prime \cos k_x \cos k_y - \mu,
    \label{Eq: 10}
\end{eqnarray}
where $\mu$ is the chemical potential that controls the electronic density. This type of electronic dispersion is general for 2D transport in strongly correlated systems and is suitable for the description of the conduction band associated with, for example, the $\mathrm{CuO_2}$ planes of high-$T_c$ cuprates. 

In the following calculations, all parameters are defined in units of $4t$ and we set $\mu/4t = -0.45$, away from the half-filled case $\mu/4t = 0.0$ (see Fig. \ref{Fig: Fermi_Surface}), since the mean-field theory yields incorrect results for a two-dimensional lattice near half-filling \cite{Micnas1990} and we avoid particle-hole symmetry \cite{Scalettar2001}. For this reason, we can take $t^\prime = 0$. We also set $V_0/4t = 1.0$, in the limit where the mean-field theory is still valid. For the summations over $\mathbf{p^\prime}$, we define $\mathbf{p^\prime} = \mathbf{k - k^\prime}$, where $\mathbf{k},\mathbf{k^\prime}$ are the momenta of the two paired electrons, which we set $|\mathbf{k}| = |\mathbf{k^\prime}| = k_F$ as two momenta in the Fermi surface. The CM momenta are then defined by fixing $\mathbf{k}$ in the direction of the point $(0, \pi)$ and by varying $\mathbf{k^\prime}$ across the Fermi surface, as shown in Fig. \ref{Fig: Fermi_Surface}.

\begin{figure}
\includegraphics[width = \linewidth]{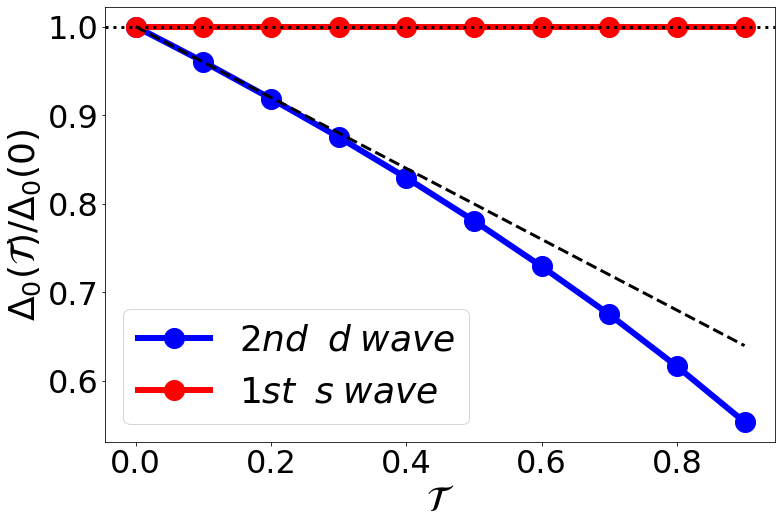}
\caption{$T = 0$ limit for the amplitude fluctuatios of the superconducting gap as a function of disorder strenght compared to the clean system. Dilute limit (red), for disorder of the 1st kind and $s-$wave symmetry, and high concentration of puddles for disorder of the 2nd kind and $d-$wave symmetry (blue). The black dashed line is a guide to the eye. Gap values are given in terms of $\Delta_0$ in the absence of disorder $\mathcal{T} = 0$.}
\label{Fig: Gaps_T0}
\end{figure}

We start by analyzing the zero temperature limit $T = 0$ of Eq. {\eqref{Eq: 9}}, where the hyperbolic tangents can be simplified. In Fig. \ref{Fig: Gaps_T0} we show how the gap amplitude $\Delta_0$ is affect by disorder $\mathcal{T}$ in the limit of disorder of the 1st type, $S(\mathbf{p^\prime}) = \delta_{0,\mathbf{p^\prime}}$, or weak concentration of puddles, and strong concentration, $S(\mathbf{p^\prime}) = 1$, in the limit of disorder of the 2nd kind. The gap amplitude is insensitive to disorder in the dilute limit for $s-$wave pairing, thus $\Delta_{0} = \Delta_{0}^{\mathrm{BCS}}$ and the BCS limit is recovered, in accordance with Anderson's theorem. However, in the opposite limit, the disorder strongly affects the amplitude of the gap for $d-$wave pairing, introducing fluctuations and decreasing its absolute value in about $50\%$ in the strong disorder limit, when compared to the clean case.

It is worth noting that the reduction is not linear as the strenght of disorder approaches the values of the fixed pairing potential, $\mathcal{T} \rightarrow V_0$, where the pertubation theory still holds. This can be traced back to the fact that the gap equation is a self-consistent equation for the aboslute value of $\Delta_0$, even after the approximations considered. Thus we see that even in the zero temperature limit, {spatially extended disorder induced by the high concentration of dopants} tends to destroy superconductivity.

\begin{figure}
\includegraphics[width = \linewidth]{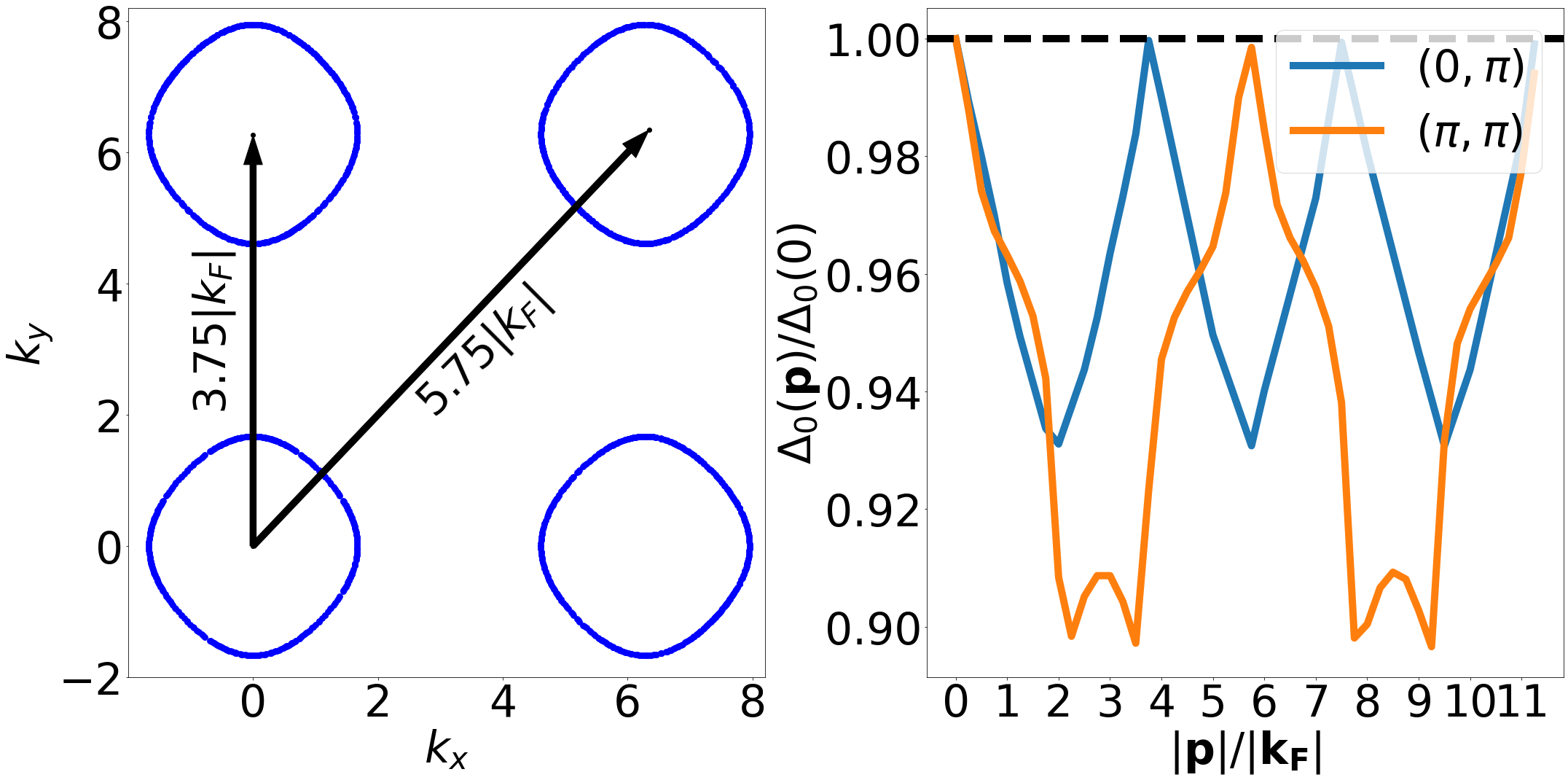}
\caption{{\it Left:} Extended Brillouin zones in the upper positive part of momentum space. The arrows indicate the distance between the centers of each Fermi surface in terms of the Fermi vector $|k_F|$ of each direction considered. {\it Right:} The amplitude of the superconducting order parameter as a function of different CM momentum vectors $|\mathbf{p}|$, in the directions $(0,\pi)$ and $(\pi, \pi)$. $\Delta_0(\mathbf{p})$ is given in units of the gap at $\mathbf{p} = 0$.}
\label{Fig: Gap_momentum}
\end{figure}

We also investigate the effects of specific finite CM momentum on the amplitude of the gap when $T = 0$. We choose a set of momenta $\{\mathbf{p}\}$ and substitute in Eq. \eqref{Eq: 9} the corresponding structure factor, namely $S(\mathbf{p_s}) = \delta_{\mathbf{p^\prime},\mathbf{p_s}}$, where $\mathbf{p_s}$ are the momenta in the set. All $\mathbf{p_s}$ are multiples of $\mathbf{k_F}$ of each direction considered, namely $(0,\pi)$ and $(\pi,\pi)$. In Fig. \ref{Fig: Gap_momentum} we display the evolution of the amplitude of the superconducting order parameter $\Delta_0$ as a function of the CM momentum of the Cooper pairs $\mathbf{p}$, for fixed disorder strenght $\mathcal{T} = 0.1$. The superconducting order parameter is modulated, with period determined by the distance between adjacents Fermi surfaces in each direction, being $3.75|\mathbf{k_F}|$ for $(0,\pi)$ and $5.75|\mathbf{k_F}|$ for $(\pi,\pi)$. Remarkably, this is in direct contact with the diffraction pattern displayed in Fig. \ref{Fig: Structure-Factor}. However, since we are considering a hard cutoff for the structure factor in terms of delta functions, the amplitude of the gap modulation is not altered by the distance from the origin. We expect that by including a more realistic model for the structure factor, the amplitudes of the modulations will decay with $\mathbf{p}$, with its effect stronger in the $(\pi,\pi)$ direction, since larger reciprocal lattice vectors $\mathbf{G}$ imply a broader structure factor, thus diminishing the amplitude of the superconducting gap. Altogether, the interplay between disorder and finite center-of-mass momentum Cooper pairs is able to strongly affect the superconducting order parameter.

\begin{figure}[!t]
\includegraphics[width = \linewidth]{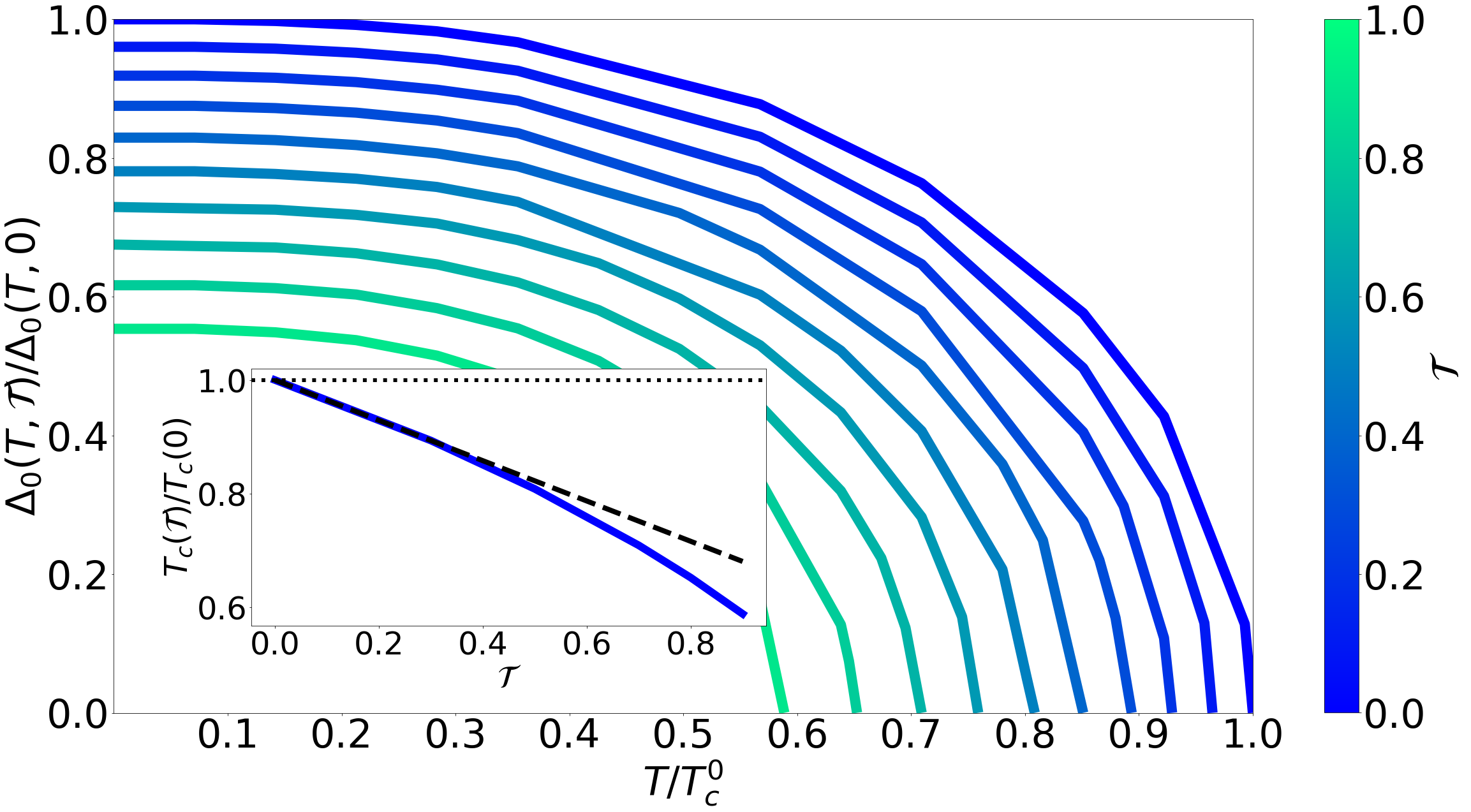}
\caption{Temperature dependence of the superconducting order parameter for different values of disorder strenght (colored bar). Gap alues are given in terms of the clean case $\mathcal{T} = 0$ and temperature in terms of $T_c^0$ also of the clean case.  {\it Inset:} The critical temperature dependence normalized to the clean value as a function of disorder strenght. The black dashed line is a guide to the eye.}
\label{Fig: Gap_T_finite}
\end{figure}

Now we turn to the finite temperature case $T\neq 0$ for the $d-$wave symmetric order parameter to understand how disorder and CM momenta for the Cooper pairs affects the critical temperature $T_c$. In Fig. \ref{Fig: Gap_T_finite} we show the evolution of the superconducting gap with temperature, for different values of the disorder strenght $\mathcal{T}$. It is clear that with increasing disorder, not only $\Delta_0(0)$ decreases, as pointed in the zero temperature limit, but we also evidence a decrease in the critical temperature $T_c$, defined as the value of temperature that $\Delta_0(T, \mathcal{T}) \rightarrow 0$, with disorder, as shown in the inset. This means that pair breaking is induced by the scattering of the finite CM momentum Cooper pairs with the nanosized puddles of the system and by increasing disorder, $T_c$ is significantly reduced. 

This pair breaking effect is due to the fact that the phase space required to pair formation is reduced when $\mathbf{p}$ increases in absolute value. In the small scattering momentum transfer sector, $\mathbf{p} < |\mathbf{k_F}|$, the gap is almost unnafacted by the presence of disorder when comprared to the value when $\mathbf{p} = 0$, since the shape of the Fermi surface intersection of the two paired electrons suffers little change. However, when $\mathbf{p}$ approaches the maximum absolute value of $2|\mathbf{k_F}|$ within the first Brillouin zone, the phase space for pair formation is greatly reduced and disorder induces pair breaking, captured by the reduction of the superconducting order paremeter. The modulation occurs for $\mathbf{p} > 2|\mathbf{k_F}|$, since electrons from different Brillouin zones participate in the scattering and pairing process. Therefore, these results point to the combined effect of finite center-of-mass momentum pairs being scattered by structural disorder induced by the network of {inhomogeneities} as a mechanism for the reduction of the superconducting gap and the critical temperature {in systems with sptially extended disorder, as in the case of networks of oxygen puddles in the overdoped cuprates or granular inhomogeneities in $\mathrm{Mo_2N_x}$.}

\section{Conclusion and discussion}
\label{Sec: 6}

{In this work we studied the effects of the presence of a spatially extended disordered granular background for doped superconducting systems, with special attention to the networks of nanosized oxygen-rich puddles in overdoped cuprates.} We show that the presence of puddles introduces strong disorder in the system that induces the formation of finite center-of-mass momentum Cooper pairs. We derive an analytical expression for the amplitude fluctuations in the superconducting gap induced by the puddles, within a mean-field BCS-like approach, in terms of the disorder strenght $\mathcal{T}$ and the finite CM momenta $\mathbf{p}$. We numerically solve this expression to show that even in the zero temperature limit the gap is strongly affected by disorder-induced CM Cooper pairs. In the limit of strong disorder, the gap tends to close and, in the finite temperature case, $T_c$ tracks the reduction of the superconducting gap {due to the presence of the spatially extended disorder in the system}. It is important to emphasize that we do not account the effect of longer-range Coulomb repulsion, restricting the application of our results to screened systems \cite{Burmistrov2012}.

The experimental observations of structural scale invariance of dopants detected by scanning micro-x-ray diffraction \cite{Poccia2010}, the promotion of critical temperature \cite{Ricci2014-1}, the agglomeration of interstitial oxygens in regions of strong local strain in the crystal structure of cuprate superconductors \cite{Song2019, Zeljkovic2014} and the proposed theoretical reports regarding the presence of networks of nanoscale superconducting islands in high-temperature superconductors \cite{Perali1996, Mello2012, Bianconi2012, Pelc2018} are in close connection with the results reported here. Eventhough we are showing that the superconducting state is depleted in the presence of strong disorder in the overdoped regime, it is clear from the above mentioned surveys that the importance of these networks and its interplay with electronic degrees of freedom pass across the whole phase diagram of hole-doped cuprates. {Notwithstanding, the effects of disorder in the superconducting state appears to depend on the way it is treated within the model. However, we emphasize that the majority of studies account to point-like disorder centers, differently from what we propose here, where the potential produced by the dopants is extend in space. If it is to be case of the cuprates, they appear in the form of the network of oxygen puddles agglomerates of dopants, where the size of each puddle ranges from multiple unit cells. On the other hand, if the system is question is not specifically a cuprate, the results reported here can be applied to understand the reduction on $T_c$ due to formation of granular regions of strong disorder, such as in $\mathrm{Mo_2N_x}$ \cite{Haberkorn2018} and Al-doped $\mathrm{MgB_2}$ \cite{Karpinski2005, Bateni2016}.}

From Ref. \cite{Tromp2023}, using scanning tunneling spectroscopy, the puddles are observed to be present in the overdoped regime of $\mathrm{(Pb, Bi)_2Sr_2CuO_{6+\delta}}$ and as the authors point out, the breakdown of superconductivity is not the consequence of a vanishing pairing interaction and does not follow the Bogoliubov-de Gennes description. From the perspective of the present work, this might be represented by the finite center-of-mass momentum of the Cooper pairs. Although the pairing interaction still present, the phase space for pair formation is diminished by $\mathbf{p}$, leading to pair breaking at a critical momentum $\mathbf{p_c}$, which can be connected to the finite gap filling observed in the experiment. However, this gap filling might be connected to some strange metal physics or collective excitations, which from the perspective of Ref. \cite{Velasco2020}, might be associated with the synchronized phases of the Cooper pairs. In summary, we point that from the experimental point of view, the puddles are remarkably important to the characterization of the superconducting state of cuprates, that should draw the attention of the theoretical community in order to deepen the understanding of the relation between the puddles and the high-T$_c$ superconductivity.

In Ref. \cite{Velasco2020} is shown how the complex networks formed by the oxygen puddles can evolve to a synchronized phase, controlled by the superfluid density, in a way that the concentration of dopant atoms controls the emergence of local superconductivity in the underdoped regime and how the systems evolves to a bulk superconductor as the concentration of dopants, thus puddles, increases as the systems approaches the optimally doped regime. It is important to emphasize that within this framework, the state studied in this work is described by the bulk superconductor state in the synchronized phase of the network formed by the oxygen puddles (see Fig. \ref{Fig: Puddles}), in the sense that we require the network of puddles to be fully synchronized in order to the band of electrons to interact with the global mode of vibration of the synchronized network. 

The approach of this work is based on a mean-field approximation for the complex network of puddles, therefore we point to the importance of considering different topologies for the organization of the puddles and how this can affect not only the transition to the superconducting state \cite{Bianconi2012-1}, but also its possible interplay with the superconducting fluctuations of pre-formed Cooper pairs observed in the pseudogap phase above $T_c$ in cuprates \cite{Dubroka2011}, as well as in conventional superconductors such as $\mathrm{NbN}$ thin films \cite{Kamlapure2013}, in terms of local superconductivity. We reinforce that our mean-field treatment is a simplified picture of any actual unconventional superconducting system, since quantum fluctuations are not explicitly included in the calculations, which are known to play a significant role for example in underdoped cuprates and disordered conventional $s$-wave superconductors \cite{Pratap2022}. Within the synchronization picture of the network of puddles, these phase fluctuations are accounted for in the picture of local formation of superconducting pairs inside the puddles, as described in Ref. \cite{Velasco2020}. Furthermore, with the increase in the superfluid density, phase rigidity is achieved in the system in the form of a synchronization transition. Nevertheless, the present work focus in understanding the effects of spatially extended disorder in the superconducting state of highly inhomogeneous systems and how this is able to generate nonzero center-of-mass momentum Cooper pairs, but a plethora of important issues regarding the physics of overdoped cuprates and strongly inhomogeneous superconductors are still unresolved.

\appendix
\section{Unitary transformation}
\label{App: A}
In this Appendix section, we show the derivation of the effective Hamiltonian containing the pairing interaction between two electrons forming a Cooper pair with finite center-of-mass momentum. The starting point is the full Hamiltonian written in momentum space $H = H_{el} + H_{p} + H_{el-p}$, which is the summation over the contributions of the electrons, puddles and electron-puddle interaction, respectively. Introducing an unitary transformation of the form $H^{\prime} = e^{-S}He^S$, where $S$ is the transformation matrix introduced in Sec. \ref{Sec: 2}, we can expand the exponentials up to second order in powers of $S$ to write the transformed Hamiltonian as 

\begin{eqnarray}
H^\prime = H + [H,S] + \frac{1}{2}[[H,S],S] + \mathcal{O}(S^3).
\end{eqnarray}
{The transformation is performed up to second order in $S$ since we suppose that the transformation matrix, the {\it ansatz} introduced in Eq. \ref{Eq: S} of the main text, is proportional to the electron-vibration coupling $g(\mathbf{k - k^\prime})$, which is the small parameter for the expansion.} By treating $H_{el-p}$ as a perturbation, we can divide the full Hamiltonian as $H = H_0 + H_{el-p}$, where $H_0$ contains the kinetic terms of electrons and puddles, to write

\begin{eqnarray}
    H^{\prime} = H_0 +  H_{el-p} + [H_0,S] + [H_{el-p},S] + \frac{1}{2}[[H_0,S],S].
    \nonumber
\end{eqnarray}
Since the goal is to eliminate the interaction, the defining equation for the transformation matrix comes from the elimination of the first-order term $[H_0, S] + H_{el-p} = 0$, from which we can extract the factors $x$ and $y$ for $S$. In this way, the transformed Hamiltonian can be written in terms of an effective interaction that comes from recombining the terms in the commutators

\begin{eqnarray}
    H^{\prime} = H_0 + \frac{1}{2}[H_{el-p},S],
\end{eqnarray}
thus the problem is reduced to an effective system described by $H = H_0 + H_{eff}$, where $H_{eff} = \frac{1}{2}\left[H_{el-p},S\right]$. By performing the calculation over the commutator $[H_0, S]$, the choice of $x$ and $y$ that eliminate the first-order term is given by

\begin{eqnarray}
    x_{\mathbf{k,k^\prime,q}} = \frac{1}{\xi_{\mathbf{k^\prime}} - \xi_{\mathbf{k}} - \omega_\mathbf{q}},\nonumber\\
    y_{\mathbf{k,k^\prime,q}} = \frac{1}{\xi_{\mathbf{k^\prime}} - \xi_{\mathbf{k}} + \omega_\mathbf{q}},\nonumber
\end{eqnarray}
and the transformation matrix $S$ is fully defined. Then we proceed to the calculation of the effective Hamiltonian that comes from the commutator of the now defined matrix $S$ and the electron-puddle interaction, which gives a combination of $M(\mathbf{q,Q})M(\mathbf{-q,\mathbf{Q^\prime}})$, where $\mathbf{Q = k - k^\prime}$ and $\mathbf{Q^\prime = k^{\prime\prime} - k^{\prime\prime\prime}}$ are two auxiliar variables that accomodate the variety of indices arising from the commutator. Recalling the definition of the factor $M$ given in the main text, we see that
\begin{eqnarray}
M(\mathbf{q,Q})M(\mathbf{-q,\mathbf{Q^\prime}}) &=& \sum_{\mathbf{R,R^\prime}}g(\mathbf{Q})g(\mathbf{Q^\prime})\nonumber\\
&\times&e^{i(\mathbf{R - R^\prime})\cdot \mathbf{q}} e^{-i(\mathbf{Q\cdot R + Q^\prime \cdot R^\prime})},\nonumber
\end{eqnarray}
which can be simplified by taking $\mathbf{R = R^\prime}$ since each $\mathbf{R}$ describes the position of a nanosized puddle and we are assuming the dilute limit of oxygen puddles, as discussed in the main text, in accordance with STEM and STM measurements \cite{Song2019,Zeljkovic2014}. In this way, the effective Hamiltonian is written as

\begin{eqnarray}
    H_{eff} &=& \sum_{\mathbf{k^\prime,k^{\prime\prime\prime},q,Q,Q^\prime}}V(\mathbf{q,Q,Q^\prime})M(\mathbf{q,Q})M(\mathbf{-q,Q^\prime})\nonumber\\
    &\times& c^{\dagger}_{\mathbf{k^{\prime\prime\prime} + Q^\prime}}c^\dagger_{\mathbf{k^\prime + Q}}c_{\mathbf{k^\prime}}c_{\mathbf{k^{\prime\prime\prime}}},
\end{eqnarray}
with $V(\mathbf{q,Q,Q^\prime}) = \omega_q/[(\xi_{\mathbf{k^{\prime\prime\prime}}} - \xi_{\mathbf{k^{\prime\prime\prime} + Q^\prime}})^2 - \omega_q^2]$. Proceeding with the calculation, we note that within BCS theory, the effective Hamiltonian describes the interaction between electrons with opposite momenta $\mathbf{k^\prime = -k^{\prime\prime\prime}}$, with zero CM momentum. However, in our case, the auxiliar variables $\mathbf{Q}$ and $\mathbf{Q^\prime}$ introduces a momentum transfer connected with a finite CM momentum for the pairs, for each fermionic operator in the effective Hamiltonian that comes from the commutator $[H_{el-p}, S]$. In this sense, we perform a change of variables introducing the finite CM momentum $\mathbf{k^\prime + k^{\prime\prime\prime} = p}$, in a way that we can eliminate the dependence on the auxiliar variables. The new variables introduced are written as $\mathbf{k = k^{\prime\prime\prime} + Q^\prime}$ and $-\mathbf{k + p^\prime = k^\prime + Q}$, where $\mathbf{p}$ and $\mathbf{p^\prime}$ are the CM momenta of the Cooper pairs. In the limit where the interaction $g(\mathbf{k,k^\prime})$ is independent of the CM momenta, we can decouple the effective interaction and end up with the effective Hamiltonian

\begin{eqnarray}
H_{\mathrm{eff}} = \sum_{\mathbf{k}, \mathbf{k}^{\prime}} \sum_{\mathbf{p}, \mathbf{p}^{\prime}} V(\mathbf{k,k^\prime})f(\mathbf{p,p^\prime})c_{\mathbf{k}, \uparrow}^{\dagger} c_{\mathbf{p}-\mathbf{k}, \downarrow}^{\dagger} c_{\mathbf{p}^{\prime}-\mathbf{k}^{\prime}, \downarrow} c_{\mathbf{k}^{\prime}, \uparrow}, \nonumber
\end{eqnarray}
with

\begin{eqnarray}
    V(\mathbf{k,k^\prime}) &=& \frac{\omega_0}{(\xi_{\mathbf{k^\prime}} - \xi_{\mathbf{k}})^2 - \omega_0^2}|g(\mathbf{k-k^\prime})|^2 \nonumber \\
    f(\mathbf{p,p^\prime}) &=& \sum_\mathbf{R} e^{-i(\mathbf{p^\prime - p})\cdot \mathbf{R}}\nonumber
\end{eqnarray}
where we assume $\omega_\mathbf{q} = \omega_0$, a dispersionless phonon mode for each puddle.


\newpage


\begin{thebibliography}{9}
\bibitem{BCS1957}
J. Bardeen, L. N. Cooper, and J. R. Schrieffer, Theory of Superconductivity. {\it Phys. Rev.} {\bf 108}, 1175 (1957)

\bibitem{Agterberg2020}
D. F. Agterberg, J. S. Davis, S. D. Edkins, E. Fradkin, D. J. Van Harlingen, S. A. Kivelson, P. A. Lee, L. Radzihovsky, J. M. Tranquada, and Y. Wang, The Physics of Pair-Density Waves: Cuprate Superconductors and Beyond. {\it Annu. Rev. Condens. Matter Phys.} {\bf 11}, 231 (2020).

\bibitem{Wang2015}
Y. Wang, D. F. Agterberg, and A. Chubukov, Coexistence of Charge-Density-Wave and Pair-Density-Wave Orders in Underdoped Cuprates. {\it Phys. Rev. Lett.} {\bf 114}, 197001 (2015).

\bibitem{Chakraborty2019}
D. Chakraborty, M. Grandadam, M. H. Hamidian, J. C. S. Davis, Y. Sidis, and C. P epin, Fractionalized pair density wave in the pseudogap phase of cuprate superconductors. {\it Phys. Rev. B} {\bf 100}, 224511 (2019).

\bibitem{Wardh2017}
J. Wardh and M. Granath, Effective model for a supercurrent in a pair-density wave. {\it Phys. Rev. B} {\bf 96}, 224503 (2017).

\bibitem{Choubey2020}
P. Choubey, S. H. Joo, K. Fujita, Z. Du, S. D. Edkins, M. H. Hamidian, H. Eisaki, S. Uchida, A. P. Mackenzie, J. Lee, J. C. S. Davis, and P. J. Hirschfeld, Atomic-scale electronic structure of the cuprate pair density wave state coexisting with superconductivity. {\it Proc. Natl. Acad. Sci. USA} {\bf 117}, 14805 (2020).

\bibitem{Loder2010}
Florian Loder, Arno P. Kampf, and Thilo Kopp, Superconducting state with a finite-momentum pairing mechanism in zero external magnetic field. {\it Phys. Rev. B} {\bf 81}, 020511(R) (2010)

\bibitem{Hamidian2016}
M. H. Hamidian, S. D. Edkins, S. H. Joo, A. Kostin, H. Eisaki, S. Uchida, M. J. Lawler, E.-A. Kim, A. P. Mackenzie, K. Fujita, J. Lee, and J. C. S. Davis, Detection of a Cooper-pair density wave in $\mathrm{Bi_2Sr_2CaCu_2O_{8+x}}$. {\it Nature} {\bf 532}, 343 (2016)

\bibitem{Liu2021}
X. Liu, Y. X. Chong, R. Sharma, and J. C. S. Davis, Discovery of a Cooper-pair density wave state in a transition-metal dichalcogenide. {\it Science} {\bf 372}, 1447 (2021).

\bibitem{Chen2021}
H. Chen {\it et al}. Roton pair density wave in a strong-coupling kagome superconductor. {\it Nature} {\bf 599}, 222 (2021).

\bibitem{Chen2018}
Angela Q. Chen, Moon Jip Park, Stephen T. Gill, Yiran Xiao, Dalmau Reig-i-Plessis, Gregory J. MacDougall, Matthew J. Gilbert and Nadya Mason, Finite momentum Cooper pairing in three-dimensional topological insulator Josephson junctions. {\it Nature Communications} {\bf 9}, 3478 (2018)

\bibitem{Edkins2019}
S. D. Edkins, A. Kostin, K. Fujita, A. P. Mackenzie, H. Eisaki, S. Uchida, S. Sachdev, M. J. Lawler, E.-A. Kim, J. C. Seamus Davis, and M. H. Hamidian, Magnetic field-induced pair density wave state in the cuprate vortex halo. {\it Science} {\bf 364}, 976 (2019)

\bibitem{Semenikhin2003}
I. A. Semenikhin, Influence of disordering on the critical temperature of superconductors with a short coherence length. {\it Physics of the Solid State} {\bf 45}, 1622 (2003)

\bibitem{Annica2022}
Debmalya Chakraborty and Annica M. Black-Schaffer, Interplay of finite-energy and finite-momentum superconducting pairing. {\it Phys. Rev. B} {\bf 106}, 024511 (2022)

\bibitem{Wen2019}
J.-J. Wen {\it et al,} Observation of two types of charge-density-wave orders in superconducting $\mathrm{La_{2-x}Sr_xCuO_4}$. {\it Nature Communications} {\bf 10}, 3269 (2019)

\bibitem{McElroy2005}
K. McElroy, H. Eisaki, S. Uchida, and S. C. Davis, Atomic-Scale Sources and Mechanism of Nanoscale Electronic Disorder in $\mathrm{Bi_2Sr_2CaCu_2O_{8+\delta}}$. {\it Science} {\bf 309}, 1048 (2005).

\bibitem{Poccia2014}
Nicola Poccia, Matthieu Chorro, Alessandro Ricci, Wei Xu, Augusto Marcelli, Gaetano Campi, Antonio Bianconi, Percolative superconductivity in $\mathrm{La_2CuO_{4.06}}$ by lattice granularity patterns with scanning micro x-ray absorption near edge structure. {\it Appl. Phys. Lett.} {\bf 104}, 221903 (2014)

\bibitem{Ricci2014}
Alessandro Ricci {\it et al}, Networks of superconducting nano-puddles in 1/8 doped $\mathrm{YBa_2Cu_3O_{6.5+y}}$ controlled by thermal manipulation. {\it New J. Phys.} {\bf 16}, 053030 (2014)

\bibitem{Huang2017}
E. W. Huang, D. J. Scalapino, T. A. Maier, B. Moritz, and T. P. Devereaux, Decrease of d-wave pairing strength in spite of the persistence of magnetic excitations in the overdoped Hubbard model. {\it Phys. Rev. B} {\bf 96}, 020503(R) (2017)

\bibitem{Balatsky2006}
A. V. Balatsky, I. Vekhter, and Jian-Xin Zhu, Impurity-induced states in conventional and unconventional superconductors. {\it Rev. Mod. Phys.} {\bf 78}, 373 (2006).

\bibitem{Rullier2008}
F. Rullier-Albenque, H. Alloul, F. Balakirev, and C. Proust, Disorder, metal-insulator crossover and phase diagram in high-Tc cuprates, {\it EPL} {\bf 81}, 37008 (2008)

\bibitem{LeeHone2020}
N. R. Lee-Hone, H. U. Ozdemir, V. Mishra, D. M. Broun, and P. J. Hirschfeld, Low energy phenomenology of the overdoped cuprates: Viability of the Landau-BCS paradigm. {\it Phys. Rev. Research} {\bf 2}, 013228 (2020) 

\bibitem{Peter2008}
Peter Henseler, Johann Kroha, and Boris Shapiro, Self-consistent study of Anderson localization in the Anderson-Hubbard model in two and three dimensions. {\it Phys. Rev. B} {\bf 78}, 235116 (2008)

\bibitem{Nguyen2022}
T. H. Y. Nguyen, D. A. Le and A. T. Hoang, Anderson localization in the Anderson–Hubbard model with site-dependent interactions. {\it New J. Phys.} {\bf 24}, 053054 (2022)

\bibitem{Nathan2021}
Nathan Giovanni, Marcello Civelli, and Maria C. O. Aguiar, Anderson localization effects on the doped Hubbard model. {\it Phys. Rev. B} {\bf 103}, 245134 (2021)

\bibitem{Anderson1959}
P. W. Anderson, Theory of Dirty Superconductors. {\it J. Phys. Chem. Solids} {\bf 11}, 26 (1959).

\bibitem{Abrikosov1958}
A. A. Abrikosov and L. P. Gor'kov, On the theory of superconducting alloys. 1. The electrodynamics of alloys at absolute zero. {\it Zh. Eksp. Teor. Fiz.} {\bf 35}, 1558 (1958).

\bibitem{Abrikosov1959}
A. A. Abrikosov and L. P. Gor'kov, Superconducting alloys at finite temperatures, {\it Zh. Eksp. Teor. Fiz.} {\bf 36}, 319 (1959). 

\bibitem{Cren2000}
T. Cren, D. Roditchev, W. Sacks, J. Klein, J.-B. Moussy, C. Deville-Cavellin, and M. Lagues, Influence of Disorder on the Local Density of States in High- Tc Superconducting Thin Films. {\it Phys. Rev. Lett.} {\bf 84}, 147 (2000)

\bibitem{Gastiasoro2018}
Maria N. Gastiasoro and Brian M. Andersen, Enhancing superconductivity by disorder, {\it Phys. Rev. B} {\bf 98}, 184510 (2018)

\bibitem{Li2021}
Zi-Xiang Li, Steven A. Kivelson and Dung-Hai Lee, Superconductor-to-metal transition in overdoped cuprates. {\it npj Quantum Materials} {\bf 6}, 36 (2021)

\bibitem{Dodaro2018}
John F. Dodaro and Steven A. Kivelson, Generalization of Anderson's Theorem for Disordered Superconductors. {\it Phys. Rev. B} {\bf 98}, 174503 (2018)

\bibitem{Campi2013}
Gaetano Campi, Alessandro Ricci, Nicola Poccia, Luisa Barba, Gianmichele Arrighetti, Manfred Burghammer, Alessandra Stella Caporale, and Antonio Bianconi, Scanning micro-x-ray diffraction unveils the distribution of oxygen chain nanoscale puddles in $\mathrm{YBa_2Cu_3O_{6.33}}$. {\it Phys. Rev. B} {\bf 87}, 014517 (2013)

\bibitem{Ricci2013}
Alessandro Ricci, Nicola Poccia, Gaetano Campi, Francesco Coneri, Alessandra Stella Caporale, Davide Innocenti, Manfred Burghammer, Martin v. Zimmermann and Antonio Bianconi, Multiscale distribution of oxygen puddles in 1/8 doped $\mathrm{YBa_2Cu_3O_{6.67}}$. {\it Scientific Reports} {\bf 3}, 2383 (2013)

\bibitem{Poccia2020}
Nicola Poccia {\it et al,} Spatially correlated incommensurate lattice modulations in an atomically thin high-temperature $\mathrm{Bi_{2.1}Sr_{1.9}CaCu_{2}O_{8+y}}$ superconductor. {\it Phys. Rev. Materials} {\bf 4}, 114007 (2020)

\bibitem{Conradson2009}
David A. Andersson, Luis Casillas, Michael I. Baskes, Juan S. Lezama, and Steven D. Conradson, Modeling of the Phase Evolution in $\mathrm{Mg_{1-x}Al_xB_2}$ $(0<x<0.5)$ and Its Experimental Signatures. J. Phys. Chem. B  {\bf 113} 11965 (2009)

\bibitem{Adhikari2022}
Rajdeep Adhikari, Bogdan Faina, Verena Ney, Julia Vorhauer, Antonia Sterrer, Andreas Ney and Alberta Bonanni, Effect of Impurity Scattering on Percolation of Bosonic Islands and Superconductivity in Fe Implanted NbN Thin Films. {\it Nanomaterials} {\bf 12}(18), 3105 (2022)

\bibitem{Jha2013}
R. Jha, J. Jyoti, and V.P.S. Awana, Impact of Gd Doping on Morphology and Superconductivity of NbN Sputtered Thin Films. {\it J Supercond Nov Magn} {\bf 26} 3069 (2013).

\bibitem{Lewellyn2020}
Nicholas A. Lewellyn, Ilana M. Percher, JJ Nelson, Javier Garcia-Barriocanal, Irina Volotsenko, Aviad Frydman, Thomas Vojta and Allen M. Goldman, Quantum Superconductor-Metal Transitions in the Presence of Quenched Disorder. {\it Journal of Superconductivity and Novel Magnetism} {\bf 33}, 183 (2020)

\bibitem{Haberkorn2018}
N. Haberkorn, S. Bengio, S. Suárez, P. D. Pérez, M. Sirena and J. Guimpel, Effect of the nitrogen-argon gas mixtures on the superconductivity properties of reactively sputtered molybdenum nitride thin films. {\it Materials Letters} {\bf 215}, 15 (2018) 


\bibitem{Slusky2001}
J.S. Slusky {\it et al}, Loss of superconductivity with the addition of Al to MgB2 and a structural transition in $\mathrm{Mg_{1-x}Al_xB_2}$. {\it Nature} {\bf 410}, 343 (2001)

\bibitem{Muller1986}
J. G. Bednorz and K. A. Muller, Possible high Tc superconductivity in the $Ba-La-Cu-O$ system. {\it Zeitschrift fur Physik B Condensed Matter} {\bf 64}, 189 (1986)

\bibitem{Zhang2013}
Gufei Zhang {\it et al}, Global and Local Superconductivity in Boron-Doped Granular Diamond. {\it Advanced Materials} {\bf 26}, 2034 (2014)


\bibitem{Campi2015}
G. Campi {\it et al,} Inhomogeneity of charge-density-wave order and quenched disorder in a high-Tc superconductor. {\it Nature} {\bf 525}, 359 (2015)




\bibitem{Poccia2010}
Michela Fratini, Nicola Poccia, Alessandro Ricci, Gaetano Campi, Manfred Burghammer, Gabriel Aeppli and Antonio Bianconi, Scale-free structural organization of oxygen interstitials in $\mathrm{La_2CuO_{4+y}}$. {\it Nature} {\bf 466}, 841 (2010)

\bibitem{Ricci2014-1}
Alessandro Ricci {\it et al,} Networks of superconducting nano-puddles in 1/8 doped $\mathrm{YBa_2Cu_3O_{6.5+y}}$ controlled by thermal manipulation. {\it New J. Phys.} {\bf 16}, 053030 (2014)

\bibitem{Velasco2020}
V. Velasco and M. B. Silva Neto, Unconventional superconductivity as a quantum Kuramoto synchronization problem in random elasto-nuclear oscillator networks. {\it J. Phys. Commun.} {\bf 5}, 015003 (2020)




\bibitem{Zhang2022}
X. Zhang, H. Zhao and J. Zhu, Visualization and control of oxygen dopant ordering in a cuprate superconductor. {\it Materials Today Physics} {\bf 23}, 100629 (2022)

\bibitem{Slezak2008}
J. A. Slezak {\it et al}, Imaging the impact on cuprate superconductivity of varying the interatomic distances within individual crystal unit cells. {\it Proc. Natl. Acad. Sci. USA } {\bf 105}(9), 3203 (2008)

\bibitem{Petrykin2000}
V.V. Petrykin, E.A. Goodilin, J. Hester, E.A. Trofimenko, M. Kakihana, N.N. Oleynikovand and Yu.D. Tretyakov, Structural disorder and superconductivity suppression in NdBa2Cu3Oz ($z \approx 7$). {\it Physica C: Superconductivity} {\bf 340}, 16 (2000)


\bibitem{Dullens2007}
R. P. A. Dullens and A. V. Petukhov, Second-type disorder in colloidal crystals. {\it EPL} {\bf 77}, 58003 (2007)

\bibitem{Hosemann1950}
R. Hosemann, Z. Phys. {\bf 128}, 1 (1950); ibid. 465 (1950).

\bibitem{Hosemann1995}
R. Hosemann and A. M. Hindeleh, J. Macromol. Sci. $-$ Phy. {\bf B34}(4), 327-356 (1995).

\bibitem{Bergmann1971}
G. Bergmann, Eliashberg Function $\alpha^2(E)F(E)$ and the Strong-Coupling Behavior of a Disordered Superconductor. {\it Phys. Rev. B} {\bf 3}, 3797 (1971)

\bibitem{MBSN2021}
M. ElMassalami and M. B. Silva Neto, Superconductivity, Fermi-liquid transport, and universal kinematic scaling relation for metallic thin films with stabilized defect complexes. {\it Phys. Rev. B} {\bf 104}, 014520 (2021)





\bibitem{Kuramoto1975}
Y. Kuramoto, Self-entrainment of a population of coupled non-linear oscillators (International Symposium on Mathematical Problems in Theoretical Physics, Lecture Notes in Physics, vol 39) ed H Araki (Berlin: Springer) 420 (1975)

\bibitem{Kuramoto1987}
Y. Kuramoto and I. Nishikawa, Statistical macrodynamics of large dynamical systems. Case of a phase transition in oscillator communities. {\it J. Stat. Phys.} {\bf 49}, 569 (1987)

\bibitem{Gogny1975}
D. Gogny, Simple separable expansions for calculating matrix elements of two-body local interactions with harmonic oscillator functions. {\it Nuclear Physica A} {\bf 237}(3), 399 (1975)

\bibitem{Song2019}
D. Song {\it et al}, Visualization of Dopant Oxygen Atoms in a $\mathrm{Bi_2Sr_2CaCu_2O_{8+\delta}}$ Superconductor. {\it Adv. Funct. Mater} {\bf 29}, 1903843 (2019)

\bibitem{Zeljkovic2014}
I. Zeljkovic {\it et al}, Nanoscale Interplay of Strain and Doping in a High-Temperature Superconductor. {\it Nano Letters} {\bf 14}(12),  6749 (2014)

\bibitem{Frohlich1937}
H. Frohlich, Theory of electrical breakdown in ionic crystals. {\it Proc. R. Soc. Lond. A} {\bf 160}(901), 230 (1937)

\bibitem{Holstein1959}
T. Holstein, Studies of polaron motion: Part I. The molecular-crystal model. {\it Annals of Physics} {\bf 8}(3), 325 (1959)

\bibitem{Fulde1964}
P. Fulde and A. Ferrell, Superconductivity in a Strong Spin-Exchange Field. {\it Phys. Rev.} {\bf 135}, A550 (1964).

\bibitem{Larkin1965}
A. I. Larkin and Yu. N. Ovchinnikov, Nonuniform State of Superconductors. {\it Sov. Phys. JETP} {\bf 20}, 762 (1965)

\bibitem{Doh2006}
Hyeonjin Doh, Matthew Song, and Hae-Young Kee, Novel Route to a Finite Center-of-Mass Momentum Pairing State for Superconductors: A Current-Driven Fulde-Ferrell-Larkin-Ovchinnikov State. {\it Phys. Rev. Lett.} {\bf 97}, 257001 (2006)

\bibitem{Woods1954}
Roger D. Woods and David S. Saxon, Diffuse Surface Optical Model for Nucleon-Nuclei Scattering. {\it Phys. Rev.} {\bf 95}, 577 (1954)

\bibitem{Gogny1973}
D. Gogny, in Proceeding of the International Conference on Nuclear Physics, Munich, edited by J. De Boer and H. J. Mang, (North-Holland, Amsterdam, 1973), Vol. 1, p. 48.

\bibitem{Gogny1975_1}
D. Gogny, in Nuclear Self-Consistent Fields, Trieste, edited by G. Ripka and M. Porneuf (North-Holland, Amsterdam, 1975), p. 333.

\bibitem{Decharge1980}
J. Decharge and D. Gogny, Hartree-Fock-Bogolyubov calculations with the D1 effective interaction on spherical nuclei. {\it Phys. Rev. C} {\bf 21}, 1568 (1980).

\bibitem{Gonzalez2018}
C. Gonzalez-Boquera, M. Centelles, X. Vinas and L. M. Robledo, New Gogny interaction suitable for astrophysical applications. {\it Physics Letters B} {\bf 779}, 195 (2018)

\bibitem{He2006}
Y. He, T. S. Nunner, P. J. Hirschfeld, and H.-P. Cheng, Local Electronic Structure of $\mathrm{Bi_2Sr_2CaCu_2O_8}$ near Oxygen Dopants: A Window on the High-Tc Pairing Mechanism. {\it Phys. Rev. Lett.} {\bf 96}, 197002 (2006)




\bibitem{Micnas1990}
R. Micnas, J. Ranninger, and S. Robaszkiewicz, Superconductivity in narrow-band systems with local nonretarded attractive interactions. {\it Rev. Mod. Phys.} {\bf 62}, 113 (1990)

\bibitem{Scalettar2001}
P. J. H. Denteneer, R. T. Scalettar and N. Trivedi, Particle-Hole Symmetry and the Effect of Disorder on the Mott-Hubbard Insulator. {\it Phys. Rev. Lett.} {\bf 87}, 146401 (2001) 

\bibitem{Burmistrov2012}
I. S. Burmistrov, I. V. Gornyi, and A. D. Mirlin, Enhancement of the Critical Temperature of Superconductors by Anderson Localization. {\it Phys. Rev. Lett.} {\bf 108}, 017002 (2012)


\bibitem{Perali1996}
A. Perali, A. Bianconi, A. Lanzara and N.L. Saini, The gap amplification at a shape resonance in a superlattice of quantum stripes: A mechanism for high-$T_c$. {\it Solid State Communications} {\bf 100}(3), 181 (1996)

\bibitem{Mello2012}
E. V. L. de Mello1, Description and connection between the oxygen order evolution and the superconducting transition in $\mathrm{La_2CuO_{4+y}}$. {\it EPL} {\bf 98}, 57008 (2012)

\bibitem{Bianconi2012}
Ginestra Bianconi, Superconductor-insulator transition on annealed complex networks. {\it Phys. Rev. E} {\bf 85}, 061113 (2012)

\bibitem{Pelc2018}
D. Pelc {\it et al}, Emergence of superconductivity in the cuprates via a universal percolation process. {\it Nat. Commun.} {\bf 9}, 4327 (2018)

\bibitem{Karpinski2005}
J. Karpinski {\it et al}, Al substitution in $\mathrm{MgB_2}$ crystals: Influence on superconducting and structural properties. {\it Phys. Rev. B} {\bf 71}, 174506 (2005)

\bibitem{Tromp2023}
Willem O. Tromp {\it et al,} Puddle formation and persistent gaps across the non-mean-field breakdown of superconductivity in overdoped $\mathrm{(Pb, Bi)_2Sr_2CuO_{6+\delta}}$. {\it Nature Materials} {\bf 22}, 703 (2023) 

\bibitem{Bateni2016}
Ali Bateni, Emre Erdem, Sergej Repp, Stefan Weber, and Mehmet Somer, Al-doped $\mathrm{MgB_2}$ materials studied using electron paramagnetic resonance and Raman spectroscopy. {\it Applied Physics Letters} {\bf 108}, 202601 (2016)

\bibitem{Bianconi2012-1}
Ginestra Bianconi, Enhancement of Tc in the superconductor–insulator phase transition on scale-free networks. {\it J. Stat. Mech.}, P07021 (2012)

\bibitem{Dubroka2011}
A. Dubroka {\it et al}. Evidence of a precursor superconducting phase at temperatures as high as 180 K in $RBa_2Cu_3O_{7-\delta}$ $(R = Y, Gd, Eu)$ superconducting crystals from infrared spectroscopy. {\it Phys. Rev. Lett.} {\bf 106}, 047006 (2011)

\bibitem{Kamlapure2013}
Anand Kamlapure, Tanmay Das, Somesh Chandra Ganguli, Jayesh B. Parmar, Somnath Bhattacharyya and Pratap Raychaudhuri, Emergence of nanoscale inhomogeneity in the superconducting state of a homogeneously disordered conventional superconductor. {\it Scientific Reports} {\bf 3}, 2979 (2013)

\bibitem{Pratap2022}
Pratap Raychaudhuri and Surajit Dutta, Phase fluctuations in conventional superconductors. {\it J. Phys.: Condens. Matter} {\bf 34}, 083001 (2022)




\end{thebibliography}
\end{document}